\input harvmac.tex
\input epsf
\noblackbox

\newcount\figno
 \figno=0
 \def\fig#1#2#3{
\par\begingroup\parindent=0pt\leftskip=1cm\rightskip=1cm\parindent=0pt
 \baselineskip=11pt
 \global\advance\figno by 1
 \midinsert
 \epsfxsize=#3
 \centerline{\epsfbox{#2}}
 \vskip 12pt
 {\bf Fig.\ \the\figno: } #1\par
 \endinsert\endgroup\par
 }
 \def\figlabel#1{\xdef#1{\the\figno}}

\def\half{{ {1\over 2} }}

\def\frac#1#2{{#1\over #2}}

\def\R{{ {\cal R} }}
\def\P{{ {\cal P} }}
\def\Q{{ {\cal Q} }}
\def\A{{ {\cal A} }}

\lref\osv{H. Ooguri, A.Strominger, C. Vafa, ``Black Hole Attractors and the Topological String'', hep-th/0405146.}
\lref\V{C. Vafa, ``Two dimensional Yang-Mills, Black Holes and Topological
Strings'', hep-th/0406058.}
\lref\AOSV{M. Aganagic, H. Ooguri, N. Saulina, C. Vafa,
``Black Holes, q-Deformed 2d Yang-Mills and
Non-perturbative Topological Strings'', hep-th/0411280, Nucl.Phys. B715 (2005) 304-348.}
\lref\ANV{M. Aganagic, A. Neitzke, C. Vafa, ``BPS Microstates and the Open Topological String Wave Function'', hep-th/0504054.}
\lref\AOO{M. Aganagic, H. Ooguri, T. Okuda, to appear. }
\lref\DFS{ D.-E. Diaconescu, B. Florea, N. Saulina, `` A vertex formalism for local ruled surfaces'',
hep-th/0505192.}
\lref\gt{D. Gross, W. Taylor, ``Two Dimensional QCD is a String Theory'',
hep-th/9301068, Nucl.Phys. B400 (1993) 181-210;
``Twists and Wilson Loops in the String Theory of Two Dimensional QCD'',
hep-th/9303046,  Nucl.Phys. B403 (1993) 395-452.
}
\lref\BP{J.~Bryan and R.~Pandharipande, ``Local Gromov-Witten
Theory of Curves'', math.AG/0411037.}
\lref\AMViii{
M.~Aganagic, A.~Klemm, M.~Marino and C.~Vafa,
``Matrix model as a mirror of Chern-Simons theory,''
JHEP {\bf 0402}, 010 (2004)
[arXiv:hep-th/0211098].
}
\lref\W{E. Witten, ``Two dimensional gauge theories revisited'',
hep-th/9204083.}
\lref\EW{E. Witten, ``Supersymmetric Yang-Mills Theory On A Four-Manifold'',
  J.Math.Phys. 35 (1994) 5101-5135, hep-th/9403195. }
\lref\BT{M. Blau, G. Thompson, ``Derivation of the Verlinde
formula from Chern-Simons theory and the $G/G$ model'', hep-th/9305010.}
\lref\JP{E. Witten, ``Quantum field theory and the Jones Polynomial'', Com. Math.Phys. 121 : 351, 1989.}
\lref\VW{C. Vafa, E. Witten, ``A strong coupling test of S-duality'',
Nucl. Phys. B 431: 3-77, 1994, hep-th/9408074.}
\lref\Rudd{R. Rudd, ``String partition function for QCD on the torus'',
hep-th/9407176}
\lref\KS{V. Kazakov, M. Staudacher, T. Wynter,
``Advances in Large N Group Theory and the Solution of Two-Dimensional $R^2$ Gravity'',
hep-th/9601153}
\lref\ghv{D.~Ghoshal and C.~Vafa,
Nucl.\ Phys.\ B { 453}, (1995) 121,
hep-th/9506122.
}
\lref\Cardoso{G. Lopez Cardoso, B. de Wit, T. Mohaupt,
``Corrections to macroscopic supersymmetric black-hole entropy'',
Phys. Lett.B 451,(309)1999, hep-th/9812082;
``Deviations from the area law for supersymmetric black holes'',
Fortsch.Phys.48, 49(2000), hep-th/9904005;
``Macroscopic entropy formulae and non-holomorphic corrections for
supersymmetric black holes'', Nucl. Phys. B 567, 87(2000), hep-th/9906094.}

\lref\IKii{A. Iqbal, A. Kashani-Poor, ``The vertex on a strip'',
hep-th/040174.}
\lref\AKMV{
M.~Aganagic, A.~Klemm, M.~Marino and C.~Vafa,
``The topological vertex,''
arXiv:hep-th/0305132.
}
\lref\AMV{
M.~Aganagic, M.~Marino and C.~Vafa,
``All loop topological string amplitudes from Chern-Simons theory,''
Commun.\ Math.\ Phys.\  {\bf 247}, 467 (2004)
[arXiv:hep-th/0206164].
}
\lref\ovc{H. Ooguri, C. Vafa,
``Two-Dimensional Black Hole and Singularities of CY Manifolds'',
Nucl.Phys. B463 (1996) 55-72, hep-th/9511164.}
\lref\INOV{A. Iqbal, N. Nekrasov, A. Okounkov, C. Vafa, ``Quantum Foam and Topological strings",
hep-th/0312022}
\lref\OPN{ D. Maulik, N. Nekrasov, A. Okounkov, R. Pandharipande,
``Gromov-Witten theory and Donaldson-Thomas theory",
math.AG/0312059 }
\lref\NV{A. Neitzke, C. Vafa,
     ``N=2 strings and the twistorial Calabi-Yau", hep-th/0402128}
\lref\NOV{N. Nekrasov, H. Ooguri, C. Vafa, `` S-duality and Topological
Strings", hep-th/0403167 }
\lref\ADKMV{
M.~Aganagic, R.~Dijkgraaf, A.~Klemm, M.~Marino and C.~Vafa,
``Topological strings and integrable hierarchies,''
arXiv:hep-th/0312085.
}
\lref\VI{C. Vafa, ``Instantons on D-branes'',
hep-th/9512078, Nucl.Phys. B463 (1996) 435-442.}

\lref\wd{H. Ooguri, C. Vafa, ``Worldsheet
derivation of Large N duality'', Nucl.Phys. B641 (2002) 3-34, hep-th/0205297.}
\lref\ov{H. Ooguri, C. Vafa, ``Knot invariants and topological
strings", Nucl.Phys. B577 (2000) 419-438, hep-th/991213.}
\lref\yoshioka{K. Yoshioka, ``Betti numbers of moduli of stable sheaves on some surfaces'',
$S$-duality and mirror symmetry (Trieste, 1995),
Nuclear Phys. B Proc. Suppl. 46 (1996), 263--268.}
\lref\nakajima{H. Nakajima, ``Instantons on ALE spaces, quiver varieties
and Kac-Moody algebras'', Tohoku Univ. preprint 1993.}
\lref\Nekrasov{N.~A.~Nekrasov,
``Seiberg-Witten prepotential from instanton counting,''
Adv.\ Theor.\ Math.\ Phys.\  {\bf 7}, 831 (2004), hep-th/0206161.
}
\lref\DV{R. Dijkgraaf, C. Vafa, ``Matrix Models, Topological Strings,
and Supersymmetric gauge theories'',
  Nucl. Phys. B644: 3-20, 2002, hep-th/0206255. }
\lref\IK{A. Iqbal, A. Kashani-Poor, ``Instanton counting and Chern-Simons theory'',
Adv. Theor. Math. Phys. 7 (2003) 459-499, hep-th/0212279.}
\lref\DMP{R. Dijkgraaf, G.Moore, R.Plesser, ``The Partition
Function of 2d string theory'', Nucl. Phys. B 394 (1993) 356-382,
hep-th/9208031.}
\lref\MSW{J. Maldacena, A. Strominger, E. Witten,
``Black Hole Entropy in M-Theory'', hep-th/9711053,
     JHEP 9712 (1997) 002}
\lref\MD{A. Dabholkar, F. Denef, G. Moore, B. Pioline,
``Exact and Asymptotic Degeneracies of Small Black Holes'',
 hep-th/0502157, JHEP 0508 (2005) 021}
\lref\MDP{A. Dabholkar, F. Denef, G. Moore, B. Pioline,
``Precision Counting of Small Black Holes'',
hep-th/0507014, JHEP 0510 (2005) 096}
\lref\Dab{A. Dabholkar, ``Exact counting of black hole microstates'',
hep-th/0409148.}
\lref\Sen{A. Sen, ``Black holes, elementary strings and holomorphic
anomaly'', hep-th/0502126; ``Black-holes and the spectrum of half-BPS
states in N=4 supersymmetric string theory'', hep-th/0504005.}
\lref\Yin{D. Shih, X. Yin, ``Exact Black Hole Degeneracies and the Topological String'',
hep-th/0508174.}
\lref\wittentwo{E. Witten, ``Two dimensional gauge theories revisited'',
hep-th/9204083.}
\lref\VafaBA{ C. Vafa,  ``Brane/anti-Brane Systems and U(N|M) Supergroup'',
hep-th/0101218.}
\lref\Estrings{ J. A. Minahan, D. Nemeschansky, C. Vafa, N. P. Warner,
``E-Strings and N=4 Topological Yang-Mills Theories''
hep-th/9802168, Nucl.Phys. B527 (1998) 581-623.}
\lref\MM{
M.~Aganagic, A.~Klemm, M.~Marino and C.~Vafa,
``Matrix model as a mirror of Chern-Simons theory,''
JHEP {\bf 0402}, 010 (2004), hep-th/0211098.}
\lref\WD{ E. Witten, ``Supersymmetric Yang-Mills Theory On A Four-Manifold'',
 hep-th/9403195, J.Math.Phys. 35 (1994) 5101-5135.}
\lref\nakajima{H. Nakajima, ``Instantons on ALE spaces, quiver varieties
and Kac-Moody algebras'', Tohoku Univ. preprint 1993.}
\Title{
  \vbox{\baselineskip12pt \hbox{hep-th/0512245}
\hbox{HUTP-05/A0050}
\hbox{UCB-PTH-05/42}
  \vskip-.5in}
}{\vbox{
  \centerline{Branes, Black Holes and Topological Strings}
\smallskip
\centerline{on}
\smallskip
\centerline{Toric Calabi-Yau Manifolds}
}}

\centerline{Mina Aganagic,$^{1}$ Daniel Jafferis,$^2$ Natalia Saulina,$^2$}
\bigskip\medskip
\centerline{$^1$ \it University of California, Berkeley, CA 94720, USA}
\vskip .03in
\centerline{$^2$ \it Jefferson Physical Laboratory,
Harvard University, Cambridge, MA 02138, USA}
\medskip
\medskip
\medskip
\vskip .5in

We develop means of computing exact degerenacies of BPS black holes on
toric Calabi-Yau manifolds. We show that the gauge theory on the D4
branes  wrapping ample divisors reduces to 2D q-deformed Yang-Mills
theory on necklaces of ${\rm \bf P}^1$'s. As explicit examples we
consider local ${\rm\bf P}^2$, ${\rm \bf P}^1 \times {\rm \bf P}^1$
and $A_k$ type ALE space times ${\rm\bf C}$.  At large $N$ the D-brane
partition function factorizes as a sum over squares of chiral blocks,
the leading one of which is the topological closed string amplitude on
the Calabi-Yau. This is in complete agreement with the recent
conjecture of Ooguri, Strominger and Vafa.

\Date{December, 2005}

\newsec{Introduction}

Recently, Strominger, Ooguri and Vafa \osv\ made a remarkable conjecture
relating four-dimensional BPS black holes in type II string theory
compactified on a Calabi-Yau manifold $X$ to the gas of topological
strings on $X$.  The conjecture states that the supersymmetric
partition function $Z_{brane}$ of the large number $N$ of D-branes
making up the
black hole, is related to the topological string partition function
$Z^{top}$ as
$$
Z_{brane} = |Z^{top}|^2,
$$
to all orders in 't Hooft $1/N$ expansion.
This provides an explicit proposal for what computes
the corrections to the macroscopic Bekenstein-Hawking entropy
of $d=4$, ${\cal N}=2$ black holes in type II string theory.
Moreover, since the partition function $Z_{brane}$ makes sense for any $N$,
this is providing the
non-perturbative completion of the topological string theory on $X$.
A non-trivial test of the conjecture requires knowing topological string partition functions at higher genus on the one hand, and on the other explicit
computation of D-brane partition functions.
Since neither are known in general, some simplifying circumstances are needed.

Evidence that this conjecture holds was provided in \V\AOSV\ in a
special class of local Calabi-Yau manifolds which are a neighborhood
of a Riemann surface $\Sigma$.  The conjecture for black holes
preserving 4 supercharges was also tested to leading order in
\MD\MDP\Yin .  The conjecture was found to have extensions to $1/2$
BPS black holes in compactifications with ${\cal N}=4$ supersymmetry
\Dab\Sen\MD\MDP . In \ANV\ the version of the conjecture for open
topological strings was formulated.

In this paper we consider black holes on local Calabi-Yau manifolds
with torus symmetries. The local geometry with the branes should be
thought of as an appropriate decompactification limit of compact ones.
While the Calabi-Yau manifold is non-compact, by considering D4-branes
which are also non-compact as in \V\AOSV , one can keep the entropy of
the black hole finite. The non-compactness of the D4 branes turns out
to also be the necessary condition to get a large black hole in four
dimensions. Because the D-branes are noncompact, different choices of
boundary conditions at infinity on the branes give rise to different
theories.  In particular, in the present setting, a given D4 brane
theory cannot be dual to topological strings on all of $X,$ but only
to the topological string on the local neighborhood of the D-brane in
$X$. This constrains the class of models that can have
non-perturbative completion in terms of D4 branes and no D6 branes,
but includes examples such as neighborhood of a shrinking ${\rm \bf
P}^2$ or ${\rm \bf P}^1 \times {\rm \bf P}^1$ in $X$.

The paper has the following organization.  In section 2 we review the
conjecture of \osv\ focusing in particular to certain subtleties that
are specific to the non-compact Calabi-Yau manifolds. We describe
brane configurations which should be dual to topological strings on
the Calabi-Yau. In section 3 we explain how to compute the
corresponding partition functions $Z_{brane}$. The D4 brane theory
turns out to be described by qYM theory on necklaces and chains of
${\rm \bf P}^1$. Where the different ${\rm\bf P}^1$'s intersect, one
gets insertions of certain observables corresponding to integrating
out bifundamental matter from the intersecting D4 branes. The qYM theory is
solvable, and corresponding amplitudes can be computed exactly.  In
section 4 we present our first example of local ${\rm \bf P}^2.$ We
show that the 't Hooft large $N$ expansion of the D-brane amplitude is
related to the topological strings on the Calabi-Yau, and moreover and
show that the version of the conjecture of \osv\ that is natural for
non-compact Calabi-Yau manifolds \AOSV\ is upheld.  In section 5 we
consider an example of local ${\rm \bf P}^1 \times {\rm \bf P}^1.$ In
section 6 we consider N D-branes on (a neighborhood of an) $A_k$ type
ALE space.  We show that at finite $N$ our results coincide with that
of H. Nakajima for Euler characteristics of moduli spaces of $U(N)$
instantons on ALE spaces, while in the large $N$ limit we find precise
agreement with the conjecture of \osv .

\newsec{Black holes on Calabi-Yau manifolds}
Consider IIA string theory compactified
on a Calabi-Yau manifold X. The effective $d=4$,  ${\cal N}=2$
supersymmetric theory
has BPS particles from D-branes wrapping holomorphic cycles
in X. We will turn off the D6 brane charge, and consider
arbitrary D0, D2 and D4 brane charges.

\subsec{D-brane theory}
Pick a basis of 2-cycles $[C^{a}] \in H_{2}(X,Z)$, and a dual basis of
4-cycles $[D_a]\in H_4(X,Z)$, $a=1,\ldots h^{1,1}(X)$,
$$
\#(D_a \cap C^{b}) = {\delta_a}^b.
$$
This determines a basis for $h^{1,1}$ $U(1)$ vector fields in four
dimensions, obtained by integrating the RR 3-form $C_3$ on the
2-cycles $C^a$.  Under these $U(1)'s$  D2 branes in class $[C] \in
H_2(X,Z)$ and D4 branes in class $[D] \in H_4(X,Z)$ carry electric
and magnetic charges $Q_{2\;a}$ and  ${Q_4}^b$ respectively:
$$
[C] = \sum_a Q_{2\, a}\; [C^a],\qquad [D] = \sum_a {Q_4}^a \;[D_a],
$$
We also specify the D0 brane charge
$Q_0$.  This couples to the one extra $U(1)$ vector multiplet which
originates from RR 1-form.

The indexed degeneracy
$$\Omega({Q_4}^a, Q_{2\,a}, Q_0)$$ of
BPS particles in spacetime
with charges $Q_0$, $Q_{2,a},$ $Q_4^a$ can be computed by counting BPS states in
the Yang-Mills theory on the
D4 brane \VI.
This is computed by the supersymmetric path integral of the four
dimensional theory on $D$ in the topological
sector with
%
%
$$
Q_0 = {1 \over 8\pi^2}\int_D tr F\wedge F,
\qquad Q_{2\,a} = {1 \over 2\pi} \int_{C_2^a} tr F.
$$
Since $D$ is curved, this theory is topologically twisted, in fact
it is the Vafa-Witten twist of the maximally supersymetric ${\cal
N} = 4$ theory on $D$.

\subsec{Gravity theory}
When the corresponding supergravity solution exists, the massive BPS particles
are black holes in 4 dimensions, with horizon area given in terms of the charges
$$
A_{BH} = \sqrt{ {1\over 3!} C_{abc}\;
{Q_4}^a {Q_4}^b {Q_4}^c |Q_0'|}
$$
where $C_{abc}$ are the triple intersection numbers of $X,$
and $Q_0'=Q_0 -\half C^{ab} Q_{2a} Q_{2b}.$\foot{$C^{ab}C_{bd}=\delta_d^a,\quad
C_{ab}=C_{abc}Q_4^c$}
The Bekenstein-Hawking formula relates this to the entropy of the black hole
$$
S_{BH} ={1\over 4} A_{BH}.
$$
For large charges, the macroscopic entropy defined by area,
was shown to agree with the microscopic one
\VI\MSW\ .
%
%
The corrections to the entropy-area relation should be suppressed
by powers in $1/A_{BH}$ (measured in plank units).

Following \Cardoso, Ooguri, Strominger
and Vafa conjectured that, just as the leading order microscopic
entropy can be computed by the classical
area of the horizon and genus zero free energy $F_0$ of A-model
topological string on $X$, the string loop corrections to the macroscopic entropy
can be computed from higher genus
topological string on $X$:
\eqn\OSV{
Z_{YM}(Q_4^a,\varphi^a,\varphi^0) = |Z^{top}(t^a,g_s)|^2
}
where
$$
Z_{YM}({Q_4}^a, \varphi^a, \varphi^{0}) =
\sum_{Q_{2 \, a},Q_0} \;
\Omega(Q_4^a,Q_{2\, a}, Q_0) \;\; exp{\bigl{(}- Q_0 \varphi^0 - Q_{2 \, a} \varphi^{a}\bigr{)}}.
$$
is the partition function of the ${\cal N}=4$ topological Yang-Mills
with insertion of
\eqn\obs{
exp{{\bigl (} - {\varphi^0\over 8\pi^2} \int tr F\wedge F -
\sum_a {\varphi^{a}\over 2\pi} \int \omega_{a}\wedge tr F  {\bigr)}}
}
where we sum over all topological sectors.\foot{
Above,
$\omega_a$ are dual to $C^a$, $\int_{C^a}  \omega_b = {\delta^a}_b$.}
The Kahler moduli of Calabi-Yau,
$$
t^a = \int_{C^a} k+i B
$$
and the topological string coupling constant $g_s$
are fixed by the attractor mechanism:
$$
t^a  = ({1\over 2} {Q_4}^a + i \varphi^a) \,g_s
$$
$$
g_s = {{4 \pi}/ \varphi^{0}}
$$
Moreover, since the loop corrections to the macroscopic entropy
are suppressed by powers of $1/N^2$ where $N \sim (C_{abc} Q_4^aQ_4^bQ_4^c)^{1/3}$ \MSW\
the duality in \OSV\ should be a large $N$ duality in the Yang Mills theory.

\subsec{D-branes for large black holes}

Evidence that the conjecture \OSV\ holds was provided in \V\AOSV\ for
a very simple class of Calabi-Yau manifolds. We show in this paper
that this extends to a broader class, provided
that the classical area of the horizon is large.
This imposes a constraint on the divisor $D$, which is what
we turn to next.

Recall that for every divisor $D$ on $X$ there is a line bundle ${\cal L}$
on $X$ and a choice of a section $s_D$ such that $D$ is the locus where this
section vanishes,
$$
s_D = 0.
$$
Different choices of the section correspond to homologous divisors
on $X$, so the choice of $[D] \in H_{4}(X,{\rm \bf Z})$ is the
choice of the first Chern-class of ${\cal L}$ (this is just
Poincare duality but the present language will be somewhat more
convenient for us) .

The classical entropy of the black hole is large when $[D]$ is deep inside
the Kahler cone of $X$, \MSW\ , i.e. $[D]$ is a ``very ample divisor''.
Then, intersection of $[D]$
with any 2-cycle class on $X$ is positive, which guarantees that
$$
C_{abc} t^a t^b t^c \gg 0.
$$
Moreover, the attractor values of the Kahler moduli are also large and positive
$$
Re(t^{a}) \gg 0.
$$
Interestingly, this coincides with the case when the corresponding
twisted ${\cal N}=4$ theory is simple. Namely, the condition that
$[D]$ is very ample is equivalent to
$$
h^{2,0}(D) >0.
$$
When this holds, \WD,\VW\ , the Vafa-Witten theory can be solved
through mass deformation.  In contrast, when this condition is
violated, the twisted ${\cal N}=4$ theory has lines of marginal
stability, where BPS states jump, and background
dependence.\foot{We thank C. Vafa for discussions which led to the
statements here.}

In the next subsection, we will give an example
of a toric Calabi-Yau manifold with configurations of
D4 branes satisfying the above condition.

\subsec{An Example}

Take $X$ to be
$$
X = O(-3) \rightarrow {\rm\bf P}^2.
$$
This is a toric Calabi-Yau which has
a $d=2$ ${\cal N}=(2,2)$ linear sigma model description in terms of
one $U(1)$ vector multiplet and 4 chiral fields $X_i$, $i=0,\ldots 3$
with charges $(-3,1,1,1)$. The Calabi-Yau $X$ is the Higgs branch of
this theory obtained by setting the D-term potential to zero,
$$
|X_1|^2+|X_2|^2+|X_3|^2= 3|X_0|^2 + r_t
$$
and modding out by the $U(1)$ gauge symmetry.
The Calabi-Yau is fibered by $\rm{T}^3$ tori, corresponding
to phases of the four $X$'s modulo $U(1)$.
Above, $r_t>0$ is the Kahler modulus of
$X$, the real part of $t = \int_{C_t} k+ i B$.
The Kahler class $[k]$ is a multiple of the integral class
$[D_t]$ which generates $H^{2}(X,Z)$,
$[k] = r_t\;[D_t]$.

Consider now divisors on $X$. A divisor in class
$$
[D] = Q \; [D_t]
$$
is given by zero locus of a homogenous polynomial in $X_i$ of
charge $Q$ in the linear sigma model:
$$D: \qquad s^{Q}_D(X_0,\ldots,X_3)=0.$$
In fact $s^{Q}_D$ is a section of a line bundle over $X$ of degree $Q [D_t]$.
A generic such divisor
breaks the $U(1)^3$ symmetry of $X$ which comes from rotating the $T^3$ fibers.
There are special divisors which preserve these symmetries, obtained by setting
$X_i$ to zero,
$$D_i: \qquad X_i=0.$$
It follows that $[D_{1,2,3}] = [D_t]$, and that $[D_0]=-3[D_t]$.
The divisor $D_0$ corresponds to the ${\rm\bf P}^2$ itself, which
is the only compact holomorphic cycle in $X$.
\bigskip
\centerline{\epsfxsize 2.5truein\epsfbox{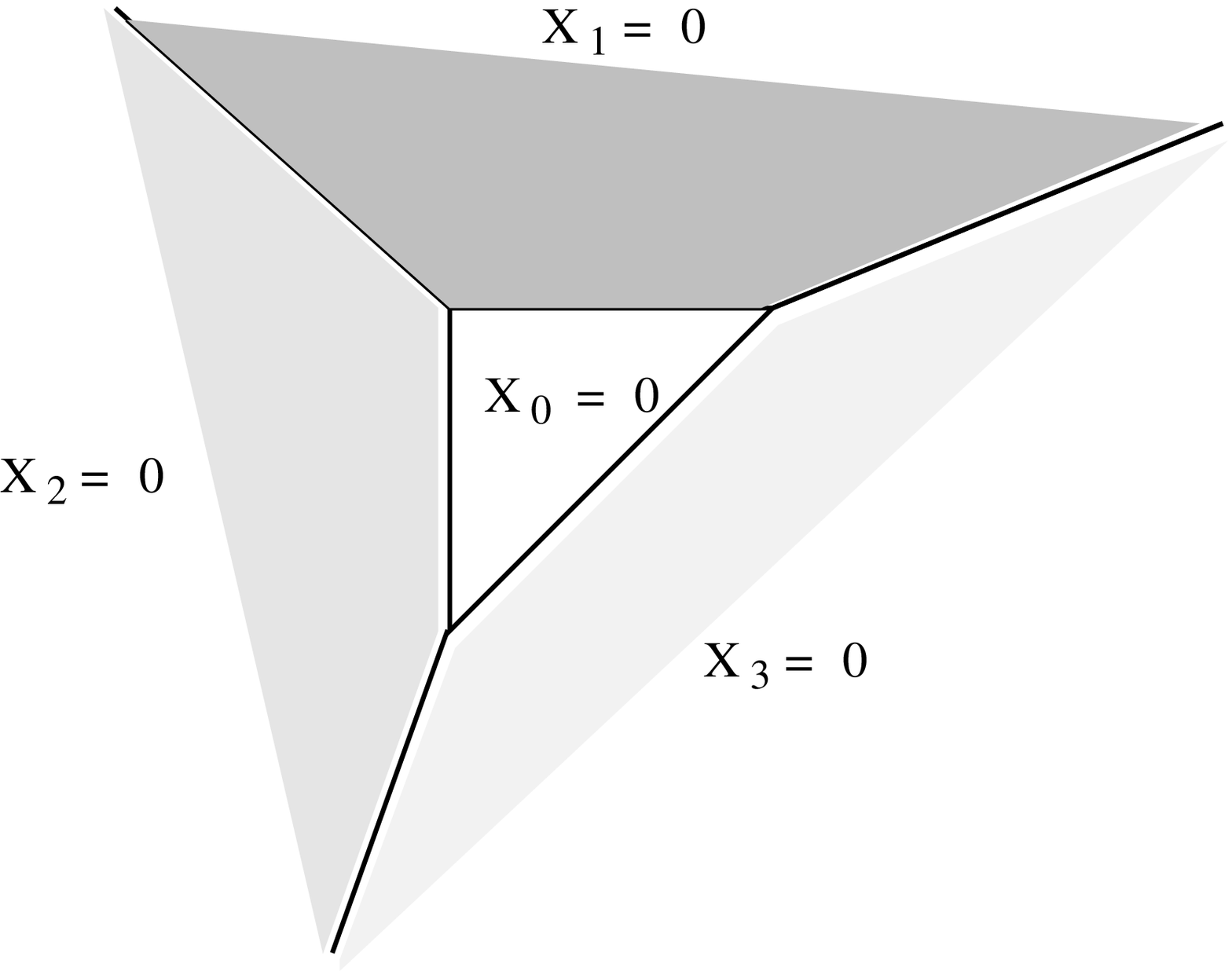}}
\leftskip 2pc
\rightskip 2pc \noindent{\ninepoint\sl
\baselineskip=2pt {\bf Fig. 1.}
{{Local ${\rm \bf P}^2.$ We depicted the base of the $T^3$ fibration which is  the interior of the convex polygon in ${\rm\bf R}^3$.
The shaded planes are its faces.
 }}}
\bigskip
As explained above, we are interested in D4 branes wrapping divisors
whose class $[D]$ is positive, $Q=Q_4>0$.
Since the compact divisors have negative classes, any divisor in
this class is non-compact 4-cycle in $X$. The divisors
have a moduli space
${\cal M}_Q$,
the moduli space of charge $Q$ polynomials,
which is very large in this case since $X$ is non-compact and the linear sigma model
contains a field $X_0$ of negative degree. If $D$ were compact, the theory
on the D4 brane would involve a sigma model on ${\cal M}_Q$.
Since $D$ is not compact, in formulating the D4 brane theory we
have to pick boundary conditions at infinity. This
picks a point in the moduli space ${\cal M}_Q$, which is
a particular divisor $D$.

Now, consider the theory on the D4 brane on $D$.  Away from the
boundaries of the moduli space ${\cal M}_Q$, the theory on the D4
brane should not depend on the choice of the divisor, but only on
the topology of $D$. In the interior of the moduli space, $D$
intersects the ${\rm\bf P}^2$ along a curve $\Sigma$ of degree
$Q$, which is generically an irreducible and smooth curve of genus
$g=(Q-1)(Q-2)/2$, and $D$ is a line bundle over it.  The theory on
the brane is a Vafa-Witten twist of the maximally supersymmetric
${\cal N}=4$ gauge theory with gauge group $G=U(1)$.  At the
boundaries of the moduli space, $\Sigma$ and $D$ can become
reducible.  For example, $\Sigma$ can collapse to a  genus zero,
degree $Q$ curve by having $s^{Q} = X_1^{Q}$, corresponding to
having $D = Q \cdot D_1.$ Then $D$ is an $O(-3)$ bundle over ${\rm
\bf P}^1$, and the theory on the D4 brane wrapping $D$ is the
twisted ${\cal N}=4$ theory with gauge group $G=U(Q)$ with scalars
valued in the normal bundle to $D$.

Both of these theories were studied recently in \AOSV\ in
precisely this context. In both cases, the theory on the D4 brane
computes the numbers of BPS bound-states of D0 and D2 brane with
the D4 brane. Correspondingly, the topological string which is
dual to this in the $1/Q$ expansion describes $only$ the maps to
$X$ which fall in the neighborhood of $D$. In other words, the D4
brane theory is computing the non-perturbative completion of the
topological string on $X_D$ where $X_D$ is the total space of the
normal bundle to $D$ in $X$. It is not surprising that the YM
theory on the (topologically) distinct divisors $D$ gives rise to
different topological string theories -- because $D$ is
non-compact, different choices of the boundary conditions on $D$
give rise to a-priori different QFTs.

It is natural to ask if there is a choice of the divisor $D$ for which
we can expect the YM theory theory to be dual to the topological
string on $X = O(-3) \rightarrow {\rm\bf P}^2$.
Consider a toric divisor in the class $[D] = Q [D_t]$
of the form
\eqn\cand{
D = N_1 D_1 + N_2 D_2 + N_3 D_3
}
where $Q = N_1+N_2+N_3$ for $N_i$
positive integers.
The D4 brane on $D$ will form bound-states with D2 branes running
around the edges of the toric base, and arbitrary number of the D0
branes.  Recall furthermore that, because $X$ has $U(1)$ symmetries,
the topological string on $X$ localizes to maps fixed under the torus
actions, i.e. maps that in the base of the Calabi-Yau project to the
edges. It is now clear that the D4 branes on $D$ in \cand\ are the
natural candidate to give the non-perturbative completion of the
topological string on $X$.  We will see in the next sections that this
expectation is indeed fully realized.

The considerations of this section suggest that of all the toric
Calabi-Yau manifolds, only a few are expected to have non-perturbative
completions in terms of D4 branes. The necessary condition translates
into having at most one compact 4-cycle in $X$,
so that the topological string on the neighborhood $X_D$ of an ample divisor
can agree with the topological string on all of $X$.
Even so, the available examples
have highly non-trivial topological string amplitudes, providing a
strong test of the conjecture.

\newsec{The D-brane partition function}

In the previous section we explained that D4-branes wrapping
non-compact, toric divisors should be dual to topological strings on
the toric Calabi-Yau threefold $X$.  The divisor $D$ in question are
invariant under $T^3$ action on $X$, and moreover generically
reducible, as the local ${\rm\bf P}^2$ case exemplifies.  In this
section we want to understand what is the theory on the D4 brane
wrapping $D$.

Consider the local ${\rm\bf P}^2$ with divisor $D$ as in \cand .
Since $D$ is reducible, the theory on the branes is a topological
${\cal N}=4$ Yang-Mills with quiver gauge group $G=U(N_1)\times U(N_2)
\times U(N_3)$.  The topology of each of the three irreducible
components is
$$
D_i : \qquad {\cal O}(-3)\rightarrow {\rm\bf P}^1
$$
In the presence of more than one divisor, there will be additional
bifundamental hypermultiplets localized along the intersections.
Here, $D_1$, $D_2$ and $D_3$ intersect pairwise along three copies of
a complex plane at $X_i=0=X_j$, $i\neq j$.

As shown in \V\AOSV , the four-dimensional twisted ${\cal N}=4$
gauge theory on
$$
{\cal O}(-p) \rightarrow {\rm \bf P}^1
$$
with \obs\ inserted is equivalent to a cousin of two dimensional
Yang-Mills theory on the base $\Sigma={\rm\bf P}^1$ with the action
\eqn\action{
S =
\frac{1}{g_s} \int_{\Sigma} {\rm tr}\, \Phi \wedge F +
\frac{\theta}{g_s} \int_{\Sigma} {\rm tr}\, \Phi\wedge \omega_{\Sigma} -
\frac{p}{2g_s}\int_{\Sigma} {\rm tr}\,\Phi^2 \wedge \omega_{\Sigma} }
where $\theta = \varphi^1/{2 \pi g_s}.$ The four dimensional theory
localizes to constant configurations along the fiber. The field
$\Phi(z)$ comes from the holonomy of the gauge field around the circle
at infinity:
\eqn\phidef{
\int_{fiber}  F(z) = \oint_{S^1_{z,\infty}} A(z)=\Phi(z).}
Here the first integral is over the fiber above a point on the base
Riemann surface with coordinate $z$.  The \action\ is the action, in
the Hamiltonian form, of a $2d$ YM theory, where
$$
\Phi(z) = g_s\, {\del\over \del A(z)}
$$
is the momentum conjugate to $A$.
However, the theory is $not$ the ordinary YM theory in two
dimensions. This is because the the field $\Phi$ is periodic. It is
periodic since it comes from the holonomy of the gauge field at
infinity. This affects the measure of the path integral for $\Phi$ is
such that not $\Phi$ but $exp(i \Phi)$ is a good variable.  The effect
of this is that the theory is a deformation of the ordinary YM theory,
the ``quantum'' YM theory \AOSV .

Integrating out the bifundamental matter fields on the intersection
should, from the two dimensional perspective, correspond to inserting
point observables where the ${\rm \bf P}^1$'s meet in the ${\rm\bf
P}^2$ base.  We will argue in the following subsections, that the
point observable corresponds to
\eqn\init{
\sum_{\R}
{\rm Tr}_{\R} \, V_{(i)}^{-1} \; {\rm Tr}_{\R}\, V_{(i+1)}
}
where
$$
V_{(i)} = e^{i \, \Phi^{(i)}- i\oint A^{(i)}}
, \qquad V_{(i+1)}= e^{i \, \Phi^{(i+1)}}
$$
The point observables $\Phi^{(i)}$ and $\Phi^{(i+1)}$ are inserted
where the ${\rm \bf P}^1$'s intersect, and the integral is around a
small loop on ${\rm\bf P}^{1}_{i}$ arount the intersection point.  The
sum is over all representations $\R$ that exist as representations of
the gauge groups on both ${\rm\bf P}^1_i$ and ${\rm\bf P}^1_{i+1}.$
This means effectively one sums over the representations of the gauge
group of smaller rank.

By topological invariance of the YM theory, the interaction \init\
depends only on the geometry near the intersections of the divisors,
and not on the global topology. For intersecting non-compact toric
divisors, this is universal, independent of either $D$ or $X$.  In the
following subsection we will derive this result.

\subsec{Intersecting D4 branes}

In this subsection we will motivate the interaction \init\ between
D4-branes on intersecting divisors.  The interaction between the D4
branes comes from the bifundamental matter at the intersection and, as
explained above, since the matter is localized and the theory
topological, integrating it out should correspond to universal
contributions to path integral over $D_L$ and $D_{R}$ that are
independent of the global geometry. Therefore, we might as well take $D$'s,
and $X$ itself to be particularly simple, and the simplest choice is
two copies of the complex 2-plane ${\rm\bf{C}}^2$ in $X={\rm\bf{C}}^3$. We can think of the pair of divisors as line bundles fibered over disks $C_a$ and
$C_b$. One might worry that something is lost by replacing $\Sigma$ by
a non-compact Riemann surface, but this is not the case -- as was
explained in \AOSV\ because the theory is topological, we can
reconstruct the theory on any $X$ from simple basic pieces by gluing,
and what we have at hand is precisely one of these building blocks.

\bigskip
\centerline{\epsfxsize 4.0truein\epsfbox{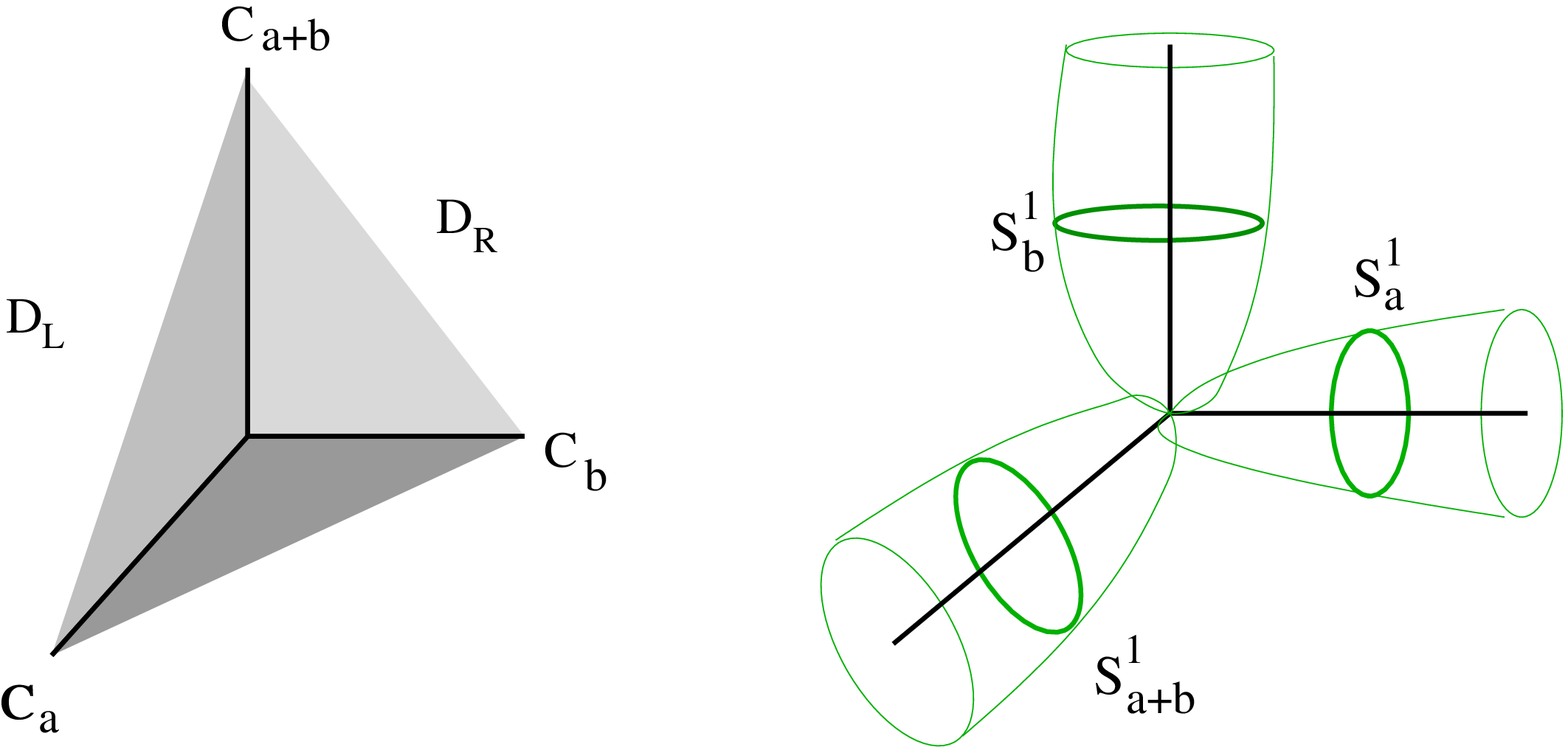}}
\leftskip 2pc
\rightskip 2pc \noindent{\ninepoint\sl
\baselineskip=2pt {\bf Fig. 2.}
{{D4-branes are wrapped on the divisors $D_{L,R}= {\rm\bf C}^2$.
The three boldfaced lines in the figure on the left correspond to three disks $C_a$, $C_b$, $C_{a+b}$ over which the $a$, $b$ and $(a+b)$ 1-cycles of the
lagrangian $T^2\times {\rm \bf R}$ fibration degenerate. The cycles of the $T^2$ which are finite are depicted in the figure on right.}}}
\bigskip

The fields at the intersection $C_{a+b}= D_L\cap D_R$
transform in the bifundamental $(M,\bar{N})$ representations of the
$U(M)\times U(N)$ gauge groups on the D-branes.
We will first argue that
the effect of integrating them out is insertion of
\eqn\inittwo{
\sum_{\R}
{\rm Tr}_{\R} \, \exp{(i \oint_{S^1_{b}} A^{(L)})} \;\, {\rm Tr}_{\R}\,
\exp{(i \oint_{S^1_{b}} A^{(R)})}
}
where
$\oint_{S^1_{b}} A^{(L)}$ and $\oint_{S^1_{b}} A^{(R)}$
are the holonomies of the gauge fields on $D_L$ and $D_R$ respectively around
the circle at infinity on the cap $C_{a+b},$
i.e. $S^{1}_{b}=\partial C_{a+b}$, see figure 2. (If this notation seems odd, it will
stop being so shortly).

We will argue this by consistency as follows.\foot{We thank C. Vafa for suggesting use of this approach.}
First, note that there is correlation between turning
of certain fluxes on $D_L$ and $D_R$. To see this note that,
if one adds D2 branes along $C_{a+b}$, the D2 branes have the effect
of turning on flux on $both$ $D_L$ and $D_R$. Consider for simplicity the case where $M=1=N$. The fact that the corresponding fluxes are correlated is the statement that
$
\int  F^{(L)} = \int F^{(R)}
$
where integrals are taken over the fibers over a point on $C_{a+b}$
in the divisors $D_L$ and $D_R$ respectively, where we view $D_{L,R}$
as fibrations over $C_{a+b}$. Since $S^{1}_{b}=\partial C_{a+b}$
this is equivalent to
\eqn\stat{
\oint_{S^{1}_{a+b}} A^{(L)} = \oint_{S^{1}_{a+b}} A^{(R)}
}
where $S^{1}_{a+b}$ is the one cycle in $X$ that $vanishes$
over $C_{a+b}$ (this cycle is well defined in $X$ as we will review shortly).
This is consistent with insertion of \eqn\oper{ \sum_{n\in {\rm\bf
Z}} \; \exp{(i{ n \oint_{S^{1}_{b}}} A^{(L)})}\; \exp{(i
{n\oint_{S^{1}_{b}} A^{(R)}})}. } because $\oint_{S^{1}_{b}}
A^{(L,R)}$ and $\oint_{S^{1}_{a+b}} A^{(L,R)}$ are canonically
conjugate, (one way to see this is to consider the qYM theory one
gets on $C_{a+b}$. Then insertion of \oper\ implies \stat\ as an
identity inside correlation functions). For general $M$, $N$ gauge
and Weyl invariance imply precisely \inittwo .

We must still translate the operators that that appear in \inittwo
, in terms of operators $\Phi^{(L,R)}$ and $A^{(L,R)}$ in the qYM
theories on $C_a$ and $C_b$.  This requires understanding of
certain aspects of $T^3$ fibrations.  While any toric Calabi-Yau
threefold is a lagrangian $T^3$ fibration, it is also a special
lagrangian $T^2\times {\rm\bf R}$ fibration, where over each of
the edges in the toric base a $(p,q)$ cycle of the $T^2$
degenerates. The one-cycle which remains finite over the edge is
ambiguous. In the case of ${\rm\bf C}^3$, we will choose a fixed
basis of finite cycles (up to $SL(2,{\rm \bf Z})$ transformations
of the $T^2$ fiber), that will make the gluing rules particularly
simple.\foot{ In the language on next subsection, this corresponds
to inserting precisely $q^{pC_{2}(\R)}$ to get ${\cal O}(-p)$ line
bundle} This is described in figure 2.  In the figure, the
1-cycles of the $T^2$ that vanish over $C_a$, $C_b$ and $C_{a+b}$
are $S^1_a$, $S^1_b$, $S^1_{a+b}$, respectively. These determine
the point observables $\Phi$'s in the qYM theories on the
corresponding disk. We have chosen a particular basis of the
1-cycles that remain finite.  From the figure it is easy to read
off that
$$
\oint_{S^1_{b}} A^{(L)} = \oint_{S^1_{a+b}} A^{(L)} - \Phi^{(L)}, \qquad \oint_{S^1_{b}} A^{(R)} = \Phi^{(R)},
$$
which justifies \init .
In the next subsection we will compute the qYM amplitudes with these observables inserted.

\subsec{Partition functions of qYM}

Like ordinary two dimensional YM theory, the qYM theory is
solvable exactly \AOSV .  In this subsection we will compute the
YM partition functions with the insertions of observables \init .
In \AOSV\ it was shown that qYM partition function $Z(\Sigma)$ on
an arbitrary Riemann surface $\Sigma$ can be computed by means of
operatorial approach. Since the theory is invariant under area
preserving diffeomorphisms, knowing the amplitudes for $\Sigma$ an
annulus $A$, a pant $P$ and a cap $C$, completely solves the
theory -- amplitudes on any $\Sigma$ can be obtained from this by
gluing.  In the present case, we will only need the cap and the
annulus amplitudes, but with insertions of observables.  Since the
Riemann surfaces in question are embedded in a Calabi-Yau, we are
effectively sewing Calabi-Yau manifolds, so one also has to keep
track of the data of the fibration. The rules of gluing a
Calabi-Yau manifold out of ${\rm \bf C}^3$ patches are explained
in \AKMV\ and we will only spell out their consequences in the
language of 2d qYM.

In the previous subsection, the theory on divisors $D_L$ and $D_R$ in
${\rm\bf C}^3$ was equivalent to qYM theories on disks $C_a$ and $C_b$, with some observable insertions.
These are Riemann surfaces with a boundary, so the corresponding path
integrals define states in the Hilbert space of qYM theory on $S^1$.
Keeping the holonomy $U =
Pe^{i \oint A}$ fixed on the boundary, the corresponding wave
function can be expressed in terms of characters of irreducible
representations ${\cal R}$ of $U(N)$ as:
$$
Z(U) = \sum_{\cal R} Z_{\cal R}\, {\rm Tr}_{\cal R} U
$$
The first thing we will answer is how to compute the corresponding states, and then we will see how to glue them together.
As we saw in the previous section, the choice of the coordinate $\oint_{S^1} A$
on the boundary is ambiguous, as the choice of the cycle which remains finite is ambiguous. This ambiguity is related to the choice of the Chern class
of a line bundle over a non-compact Riemann surface, i.e. how the divisors $D_{L,R}$ are fibered over the corresponding disks. The simplest choice is the one
that gives trivial fibration, and this is the one we made in figure 2 (this corresponds to picking the cycle that vanishes over $C_{a+b}$).

The partition function on a disk with trivial bundle over it and no insertions is
\eqn\disk{
Z(C)(U) = \sum_{\R \in U(N)} \;S_{0 \R}\; \;e^{i \theta C_1(\R)}
\;Tr_\R U,
}
Above, $C_{1}(\R)$ is the first casimir of the representation $\R$, and
$S_{\R\P}(N,g_s)$ is a relative of the S-matrix of the $U(N)$ WZW model
\eqn\defsr{
S_{\R \Q}(N,g_s)=\sum_{w\in S_N} \epsilon(w)
q^{-(\R+\rho_N) \cdot w(\Q+\rho_N)},\quad
}
where
$$
q=exp(-g_s)
$$
and $S_N$ is Weyl group of $U(N)$ and $\rho_N$
is the Weyl vector.\foot{The normalization of the path integral  is ambiguous. In our examples in sections 4-6  we will 
choose it in such a way that the amplitudes agree
with the topological string in the large $N$ limit.}

Sewing $\Sigma_L$ and $\Sigma_R$ is done by
$$
Z(\Sigma_L \cup \Sigma_R) = \int dU \;Z(\Sigma_L)(U) \;
Z(\Sigma_R)(U^{-1}) = \sum_{\cal R} Z_{\cal R}(\Sigma_L)Z_{\cal R}(\Sigma_R)
$$
For example, the amplitude corresponding to
$\Sigma={\rm \bf P}^1$ with $O(-p)$
bundle over it and no insertions can be obtained by gluing two disks
and an annulus with $O(-p)$ bundle over it:
\eqn\ann{
Z(A,p)(U_1,U_2) =\sum_{\R \in U(N)} \;q^{pC_2(\R)/2} \;e^{i \theta C_1(\R)} \;Tr_\R U_1 \;Tr_\R U_2
}
This gives
\eqn\sphere{
Z(P^1, p) =  \sum_{\cal R} (S_{0\R})^2 q^{p C_2(\R)/2} e^{i \theta C_1(\R)}
}

In addition we will need  to know how to compute expectation values of
observables in this theory.
As we will show in the appendix B, the amplitude on a cap with a trivial line bundle and observable
${\rm Tr}_Q \, e^{i\Phi-i n\oint_{S^1} A}$ inserted equals
%
%
%
\eqn\pointtwo{
Z(C, \, {\rm Tr}_{\Q}\; e^{i\Phi-i n\oint_{S^1} A}) (U)= \sum_{\R} q^{{n\over 2} C_2(\Q)}
S_{{\bar \Q} \R}(N,g_s) {\rm Tr}_{\R}U.
}
where $U$ is the holonomy on the boundary.

It remains to compute the expectation value of the observables in \init\
in the two-dimensional theory on ${C}_a$ and ${C}_b$.
The amplitude on the intersecting divisors $D_L$, $D_R$ is
\eqn\intergen{\eqalign{
Z(V)(U^{(L)}, U^{(R)}) =&
\sum_{\Q \in U(M), \P \in U(N)} \,
V_{\Q \P}(M,N)
 Tr_{\Q} U^{(L)}Tr_{\P} U^{(R)} \cr
V_{\Q \P}(M,N) = & \sum_{\R \in U(M)} \,S_{\Q {\bar \R} }(M,g_s)\, q^{\half C_2^{(M)}(\R)} \, S_{{\R} \P}(N,g_s)\,
}}
In the above, $U^{(L,R)}$
is the holonomy at  the boundary of $C_a$ and $C_b$.

When $M=N$, there is a simpler expression for the
vertex amplitude in \intergen.
Using the definition of $S_{\P \R}$ \defsr\ and summing over $\R$
we have
\eqn\transf{\eqalign{
V_{\P \Q}=
\theta^N(q) \,\, q^{-{\half C_2(\P)}}\, \,
{S_{{\overline \P} \Q}} q^{-{\half C_2(\Q)}}\, \,
}}
and where $\theta(q)=\sum_{m \in {\bf Z}} q^{m^2 \over 2} .$
This is related to the familiar realization in WZW models
of the relation
$$
STS=(TS T)^{-1}
$$
between
$SL(2,{\bf Z})$ generators $S$ and $T$ in WZW models
where
\eqn\TS{
T_{\R\Q}=q^{\half C_2(\R)}\delta_{\R\Q},\quad
S^{-1}_{\R\P}(g_s,N)=S_{\R\P}(-g_s,N)=S_{{\bar \R}\P}(g_s,N).
}
The difference is that there is no quantization of the level $k$ here.
Even at a non-integer level, this is more straightforward
in the $SU(N)$ case,
where the theta function in \transf\ would not have appeared.
\subsec{Modular transformations}

The partititon functions of D4 branes on various
divisors with chemical potentials
$$
S_{4d}=\frac{1}{2g_s} \int  \tr F \wedge F +
\frac{\theta}{g_s} \int \tr F \wedge \omega ,
$$
turned on, are computing degeneracies of bound-states
of $Q_2$ D2 branes and $Q_0$ D0 branes with the D4 branes,
where
\eqn\topclasses{Q_0 = \frac{1}{8\pi^2}\int \tr F\wedge F, ~~
Q_2=\frac{1}{2\pi} \int  \tr F \wedge \omega,}
so the YM amplitudes should have an expansion of the form
\eqn\vafawitten{
Z^{q{\rm YM}} = \sum_{q_0, q_1} \Omega(Q_0, Q_2, Q_4) \exp\left[
-{4\pi^2 \over g_s} Q_0 - {2\pi \theta \over g_s} Q_2 \right].}
The amplitudes we have given are not expansions in $exp(-1/g_s)$,
but rather in $exp(-g_s)$, so the existence of the \vafawitten\
expansion is not apparent at all. The underlying ${\cal N}=4$
theory however has $S$ duality that relates strong and weak
coupling expansions, so we should be able to make contact with
\vafawitten .

Since amplitudes on more complicated manifolds are obtained from the
simpler ones by gluing, it will suffice for us to show this for
the propagators, vertices and caps. Consider the annulus amplitude \ann\
Using the Weyl-denominator form of the $U(N)$ characters
$
Tr_\R U = \Delta_H(u)^{-1} \sum_{w\in S_N} (-)^{\omega}
e^{\omega(iu)\cdot (\R+\rho_N)}
$
we can rewrite $Z(A,p)$ as
$$
Z(A,p)(U,V) =\Delta_H(u)^{-1}\Delta_H(v)^{-1}
\sum_{n\in {\bf Z}^N} \sum_{w\in S_N} q^{{p\over 2}n^2}
e^{n(i u -w(iv))}
$$
which is manifestly a modular form,\foot{Recall,
$
\theta(\tau,u) = (-i \tau)^{-{1\over 2}}e^{-i \pi {u^2\over \tau}}
\theta(-{1\over \tau}, {u\over \tau}),
$
where $\theta(\tau,u)=\sum_{n\in {\bf Z}} e^{i \tau n^2} e^{2\pi i u}$.}
which we can write
\eqn\modanul{
Z(A,p)(U,V) =\Delta_H(u)^{-1}\Delta_H(v)^{-1} \left({g_s p \over 2\pi}\right)^{-{N\over 2}}
\sum_{m\in {\bf Z}^N} \sum_{w\in S_N} {\tilde q}^{{1\over 2 p}\Bigr(m -
{u-w(v)\over 2\pi} \Bigl)^2}
}
where in terms of ${\tilde q} = e^{-4 \pi^2/g_s}$.
In the above, the eigenvalues $U_i$ of $U$ are written as
$U_i = exp(iu_i)$, and $\Delta_H(u)$
enters the Haar measure:
$$
\int dU = \int \prod_i du_i \Delta_H(u)^2
$$
Note that, in gluing, the determinant $\Delta_H(u)^2$
factors cancel out, and simple degeneracies will be left over.

Similarly, the vertex amplitude \intergen\ corresponding to intersection of $N$ and $M$ D4 branes
can be written as (see appendix C for details):
%
%
%
%
\eqn\modul{
Z(U,V)=
\Delta_H(u)^{-1} \Delta_H(v)^{-1} \theta^M(q)
\sum_{m\in {\bf Z}^M} q^{-{1\over 2}m^2} e^{m\cdot v}
\sum_{w \in S_N}(-)^w \sum_{n \in {\bf Z}^N}
e^{n(\cdot w(i u)+ i v- g_s(\rho_N- \rho_M))}
}
where $v,\rho_M$ are regarded as $N$ dimensional vectors,
the last $N-M$ of whose entries are zero.
We see that $Z(U,V)$ is given in terms of theta functions, so it is
modular form, its modular transform given by
\eqn\modtrans{\eqalign{
Z(U,V)= \Delta_H(u)^{-1} \Delta_H(v)^{-1} \;\left({g_s\over 2\pi}\right)^{-{M/2}} \;\theta^M({\tilde q})
&\sum_{m\in {\bf Z}^M} \tilde{q}^{-{1\over 2}(m+i v/2\pi)^2}\cr
&\sum_{w \in S_N}(-)^w \sum_{n \in {\bf Z}^N}
e^{n(\cdot w(i u)+i v- g_s(\rho_N- \rho_M))}}
}
In a given problem, it is often easier to compute the degeneracies of the BPS states from the amplitude as a whole, rather than from the gluing the
S-dual amplitudes as in \modtrans . Nevertheless, modularity at the level of
vertices, propagators and caps, demonstrates
that the $1/g_s$ expansion of our amplitudes does exist in a general case.

\newsec{Branes and black holes on local ${\rm \bf P}^2.$ }

We will now use the results of the previous section to study black holes
on $X=O(-3) \rightarrow {\bf P}^2$.
As explained in section $2$, to get large black holes on ${\rm \bf R}^{3,1}$
we need to consider D4 branes wrapping very-ample divisors on $X$,
which are then necessarily non-compact. Moreover, the choice of divisor $D$
that should give rise to a dual of topological strings on $X$ corresponds to
$$
D = N_1 D_1+ N_2 D_2 + N_3 D_3
$$
where $D_i$, $i=1,2,3$ are the toric divisors of section 2.


Using the results of section $3$, it is easy to compute the amplitudes
corresponding to the brane configuration.
We have $N_1\geq N_2 \geq N_3$ D4 branes on three divisors of topology
$D_i = O(-3)\rightarrow {\bf P}^1$. From each, we get a copy of quantum Yang Mills theory on ${\rm \bf P}^1$ with $p=3$, as discussed in section $3$.
From the matter at the intersections, we get in addition,
insertion of observables \init\ at two points in each ${\rm \bf P}^1$.

\bigskip
\centerline{\epsfxsize 3.0truein\epsfbox{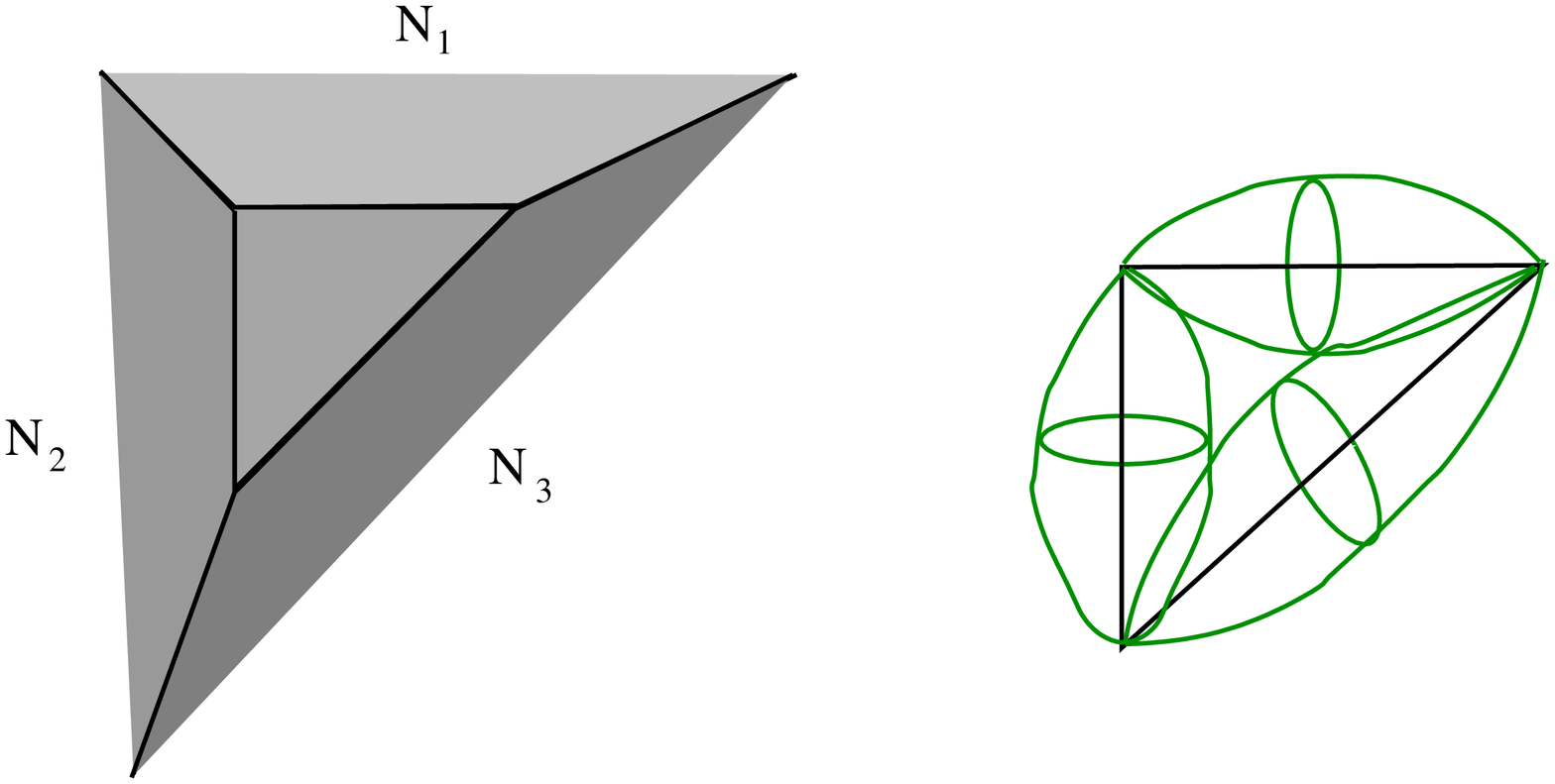}}
\leftskip 2pc
\rightskip 2pc \noindent{\ninepoint\sl
\baselineskip=2pt {\bf Fig. 3.}
{{Local $\bf P^2,$ depicted as a toric web diagram. The numbers of D4 branes wrapping the torus invariant non-compact 4-cycles are specified.
 }}}
\bigskip

All together this gives:
\eqn\zymP{
Z_{qYM}=\alpha \sum_{\R_i\in U(N_i)}
V_{\R_2 \R_1}(N_2,N_1)\, V_{\R_3 \R_2}(N_3,N_2)\, V_{\R_3 \R_1}(N_3,N_1)\,\,
\prod_{j=1}^3 q^{3C_2(\R_i) \over 2}
e^{i\theta_i C_1(\R_i)}
}
Note that in the physical theory there should be only one chemical
potential for D2-branes, corresponding to the fact that $H_2(X,{\rm
\bf Z})$ is one dimensional.  In the theory of the D4 brane we
$H_2(D,{\rm\bf Z})$ is three dimensional, generated by the 3 ${\rm \bf
P}^1$'s in $D$ -- the three chemical potentials $\theta_i$ above
couple to the D2 branes wrapping these.  While all of these D2 branes
should correspond to BPS states in the Yang-Mills theory, not all of
them should correspond to BPS states once the theory is embedded in
the string theory. Because the three ${\rm\bf P}^1$'s that the D2
brane wrap are all homologous in $H_{2}(X,{\rm \bf Z})$,
$$ [{\rm \bf
P}_1^{1}] -[{\rm \bf P}_3^{1}] \sim 0,\qquad [{\rm \bf P}_2^{1}]
-[{\rm \bf P}_3^{1}] \sim 0
$$
there will be D2 brane instantons that
can cause those BPS states that carry charges in $H_2(D,{\rm\bf
Z})$ to pair up into long multiplets.  Decomposing $H_{2}(D,{\rm
\bf Z})$ into a $H_{2}(D,{\rm \bf Z})^{||} = H_{2}(X,{\rm \bf Z})$ and
$H_{2}(D,{\rm \bf Z})^{\perp}$, it is natural to turn off the the
chemical potentials for states with charges in $H_{2}(D,{\rm \bf
Z})^{\perp}$. This corresponds to putting
$$
\theta_i=\theta, \qquad i=1,2,3.
$$
For some part, we will keep the $\theta$-angles different, but there
is only one $\theta$ natural in the theory.

The normalization $\alpha$ of the path integral is
chosen in such a way that $Z_{qYM}$ has chiral/anti-chiral
factorization in the large $N_i$ limit (see 4.6 and 4.10 below).
$$
\alpha=q^{-{(\rho^2_{N_2}+{N_2 \over 24})}}
q^{-2{(\rho^2_{N_3}+{N_3 \over 24})}}\,\,
e^{(N_1+N_2+N_3)\theta^2\over 6g_s} q^{(N_1+N_2+N_3)^3\over 72}
$$

The partition function simplifies significantly
if we take equal numbers of the
D4 branes on each $D_i$,
$$
N_i=N, \quad i=1,2,3
$$
since in this case, we can
replace \intergen\ form of the vertex amplitude with the simpler
\transf ,  and the
D-brane partition function becomes

\eqn\zymii{
Z_{qYM}=\alpha \,\theta^{3N}(q)\,\,\sum_{\R_1,\R_2,\R_3 \in U(N)}
S_{\R_1 {\bar \R_2}}(g_s,N)\, S_{\R_2 {\bar \R_3}}(g_s,N)\, S_{\R_3 {\bar \R_1}}(g_s,N)\,\,
\prod_{j=1}^3 q^{C_2(\R_i) \over 2}
e^{i\theta_i C_1(\R_i)}
}
%
%
%

In the following subsections we will first take the large $N_i$ limit of $Z_{qYM}$ to get the closed
string dual of the system. We will then use modular properties of the
partition function to compute the degeneracies of the BPS states of
D0-D2-D4 branes.

\subsec{Black holes from local ${\rm\bf P}^2$}

According to the conjecture of \osv\ (or more precisely, its
version for the non-compact Calabi-Yau manifolds proposed in \AOSV
) the large $N$ limit of the D-brane partition function
$Z_{brane}$, which in our case equals $Z_{qYM},$ should be given
by
$$
Z_{qYM}(D, g_s, \theta) \approx \sum_{\alpha} |Z_{\alpha}^{top}(t,g_s)|^2
$$
where
$$
t = {1\over 2}(N_1+N_2+N_3)g_s - i \theta
$$
since $ [D] = (N_1+N_2+N_3) [D_t]$ where $[D_t]$ is dual to the class that
generates $H_2(X,{\rm\bf Z})$. In the above, the two expressions
should equal up to terms of order $O(exp(-1/g_s))$, hence the ``approximate''
sign.  The sum over $\alpha$ is the sum over chiral blocks which
should correspond to the boundary conditions at infinity of $X$. More
precisely, the leading chiral block should correspond to including
only the normalizable modes of topological string on $X$, which count
holomorphic maps to ${\rm \bf P}^2$, the higher ones containing
fluctuation in the normal direction \AOSV\ANV .  We will see below
that this prediction is realized precisely.

The Hilbert space of the qYM theory, spanned by states labeled by
representations ${\cal R}$ of $U(N),$ at large $N$ splits into
$$
{\cal H}^{qYM} \approx \oplus_{\ell} \; {\cal H}_{\ell}^{+} \otimes {\cal H}_{\ell}^{-}
$$
where ${\cal H}_l^{+}$ and ${\cal H}_l^{-}$ are spanned by
representations $R_+$ and $R_-$ with small numbers of boxes as
compared to $N$, and ${\ell}$ is the $U(1)$ charge.  Correspondingly,
the qYM partition function also splits as
$$
Z_{qYM} \approx \sum_{\ell}\;Z_{\ell}^{+} \; Z_{\ell}^{-},
$$
where $Z_{\ell}^{\pm}$ are the chiral and anti-chiral partitions.
We will now compute these, and show that they
are given by topological string amplitudes.
\vskip 0.5cm
$i.\;$ $The$ $N_i=N$ $case.$
\vskip 0.5cm
We'll now compute the large $N$ limit of the D-brane
partition function \zymii\ for  $N_i=N$, $i=1,2,3$.
At large $N$, the $U(N)$ Casimirs in representation ${\cal R} = R_+{\bar R}_-[\ell_R]$
are given by
\eqn\kas{\eqalign{
C_2(\R)&=\kappa_{R+} + \kappa_{R-} + N(|R_+|+|R_-|)+ N{\ell_R}^2 + 2 {{\ell}_R}(|R_+|-|R_-|),\cr
C_1(\R)&=N{\ell}_R + |R_+|-|R_-|
}}
where
$$
\kappa_R=\sum_{i=1}^{N-1} R_i(R_i-2i+1)
$$
and $|R|$ is the number of boxes in $R$.

The S-matrix $S_{\R \Q}$ is at large $N$ given in \ANV\
\eqn\smatr{\eqalign{
q^{-(\rho^2 +{N\over 24}) } S_{{\cal R} {\cal Q}}(-g_s,N)=&
M(q^{-1})\eta(q^{-1})^N (-)^{\vert R_+ \vert + \vert R_- \vert
+\vert Q_+ \vert + \vert Q_- \vert}\cr
\times& q^{N {\ell}_R {\ell}_Q} q^{{\ell}_Q(\vert R_+ \vert -\vert R_-\vert)} q^{{\ell}_R(\vert Q_+ \vert -\vert Q_-\vert)}
q^{{N(\vert R_+ \vert+\vert R_- \vert+\vert Q_+ \vert+
\vert Q_- \vert)\over 2}}\cr
\times&
q^{{\kappa_{R_+}+\kappa_{R_-} \over 2}}
\sum_{P} q^{-N\vert P \vert} (-)^{\vert P \vert}
{\hat{C}}_{Q_+^T R_+ P}(q) {\hat{C}}_{Q_-^T R_- P^T}(q).
}
}
The amplitude ${\hat C}_{R P Q}(q)$ is
the topological vertex amplitude of \AKMV .\foot{The
conventions of this paper and \AKMV\ differ, as here $q= e^{-g_s}$, but $q_{there}=e^{g_s}$, consequently the topological vertex amplitude
$C_{R P Q}$ of \AKMV\ is related to the present one by
$
{\hat C}_{R P Q}(q)=C_{R P Q}(q^{-1}).
$}
%
In \smatr\  $M(q)$ and $\eta(q)$ are MacMahon and Dedekind functions.

Putting this all together, let us now parameterize the integers
${\ell}_{R_i}$ as follows
$$
3 \ell = {\ell}_{R_1}+\ell_{R_2} + \ell_{R_3},\quad 3n ={\ell}_{R_1}-{\ell}_{R_3},
\quad 3k = {\ell}_{R_2}-{\ell}_{R_3}.
$$
It is easy to see that the sum over $n$ and $k$ gives delta functions:
at large $N$
\eqn\result{
Z_{qYM}(\theta_i, g_s)\sim \delta \bigl(\,N (\theta_1-\theta_3)\,\bigr) \; \delta \bigl(\,N (\theta_2-\theta_3)\,\bigr)
\times \;Z_{qYM}^{finite}(\theta, g_s)
}
 where $\theta_i=\theta$ in the finite piece.  As we will show in
Sec. 4.2 there is the same $\delta$-function singularity as in the partition function of
the bound-states of $N$ D4 branes. There it will be clear that it
comes from summing over D2 branes with charges in $H_2(X,{\rm\bf
D})^{\perp}$, as mentioned at the beginning of this section.  The
finite piece in \result\ is given by
\eqn\partiii{
Z_{qYM}^{finite}(N, \theta, g_s) =
\sum_{m \in {\rm\bf Z}} \sum_{P_1,P_2,P_3}(-)^{\sum_{i=1}^3\vert P_i \vert}
Z^+_{P_1,P_2,P_3}\bigl(t+m g_s\bigr )
Z^+_{P_1^T,P_2^T,P_3^T}\bigl( {\bar t}-m g_s\bigr).
}
The chiral block in \partiii\ is the topological string amplitude on
$X=O(-3)\rightarrow {\rm\bf P}^2$,
\eqn\chiriii{
 Z^+_{P_1,P_2,P_3}(t)={\hat Z}_0(g_s,t)  e^{- t_0\sum_i\vert P_i \vert }
\sum_{R_1,R_2,R_3}\,
 e^{- t\sum_i\vert R_i \vert }
q^{\sum_i \kappa_{R_i}}
{\hat C}_{R_2^T R_1 P_1^T}(q)\, {\hat C}_{  R_3^T R_2 P_2^T}(q)\,
{\hat C}_{R_1^T R_3 P_3^T}(q)
}
where $t_0 = -\half {Ng_s}$
and the Kahler modulus $t$ is (we will return to the meaning of $t_0$ shortly):
$$
t={3Ng_s \over 2}-i\theta.
$$
More precisely, the chiral block with trivial ghosts $P_i=0$,
$$
Z^{+}_{0,0,0}(t,g_s) = Z^{top}(t,g_s)
$$
is exactly equal to the perturbative
closed topological string partition function for
$X= O(-3) \rightarrow {\rm\bf P}^2,$ as given in \AKMV .
This exactly agrees with the prediction of \osv .

The prefactor ${\hat Z}_0(g_s,t)$  is given by
$$ {\hat Z}_0(g_s,t)=e^{-{t^3 \over 18 g_s^2}}\,\, M^3(q^{-1})
\eta^{t\over g_s}(q^{-1})\theta^{t\over g_s}(q)
$$
As explained in \AOSV\ the factor
$\eta^{t\over g_s} \sim\eta^{{3N\over 2}}$ comes from bound
states of D0 and D4 branes \VW\ without any D2 brane charge,
and moreover, it has only genus zero contribution perturbatively.
$$
\eta^{t\over g_s} \sim exp\left(-{\pi^2 t\over 6 g_s^2}\right)
+ (non-perturbative)
$$
The factor $\theta^{t\over g_s}$ comes from the bound states of
D4 branes with D2 branes along each of three the non-compact toric legs in the normal direction to the ${\rm \bf P}^2$,
and without any D0 branes. This gives no perturbative contributions
$$
\theta^{t\over g_s} \sim 1
+ (non-perturbative)
$$
The subleading chiral blocks correspond to open topological string
amplitudes in $X$ with D-branes along the fiber direction to the ${\rm
\bf P}^2$, which can be computed using the topological vertex
formalizm \AKMV .  The appearance of D-branes was explained in \ANV\ where
they were interpreted as non-normalizable modes of the topological
string amplitudes on $X$.  The reinterpretation in terms of
non-normalizable modes of the topological string theory is a
consequence of the open-closed topological string duality on \ADKMV
. While this is a duality in the topological string theory, in the
physical string theory the open and closed string theory are the same
only provided we turn on Ramond-Ramond fluxes.  We cannot do this here
however, since this would break supersymmetry, and the only correct
interpretation is the closed string one.

To make contact with this, define
$$
Z^{+}(U_1,U_2,U_3) = \sum_{R_1,R_2,R_3} Z^{+}_{R_1,R_2,R_3}
\;Tr_{R_1}U_1\;Tr_{R_2}U_2\;Tr_{R_3}U_3.
$$
where $U_i$ are unitary matrices.  This could be viewed as an open
topological string amplitude with D-branes, or more physically, as
the topological string amplitude, with non-normalizable
deformations turned on. These are not most general
non-normalizable deformations on $X$, but only those that preserve
torus symmetries -- correspondingly they are localized along the
non-compact toric legs, just like the topological D-branes that
are dual to them are.  The non-normalizable modes of the geometry
can be identified with \ADKMV\
$$
\tau^n_{i} = g_s tr(U_i^n)
$$
where the trace is  in the fundamental representation.
We can then write \partiii\ as
$$
Z^{finite} \sim \int dU_1\, dU_2\, dU_3 \;|Z^{+}(U_1,U_2,U_3)|^2
$$
where we integrate over $unitary$
matrices
$provided$ we shift
$$
U \rightarrow U e^{-t_0}
$$
where $t_0= -{1\over 2} N g_s$. This shift is the attractor
mechanism for the non-normalizable modes of the geometry \ANV . In terms
of the natural variables ${t^n}_i$, related by $\tau^n_i=exp(-t^n_i)$
to $\tau$'s we have
\eqn\qppl{
{t^n_i} = nt_0
}
This comes about as follows \ANV . First note that size of any 2-cycle
$C$ in the geometry should be fixed by the attractor mechanism to
equal its intersection with the 4-cycle class $[D]$ of the D4 branes, in this
case $[D] = 3N [D_t]$.
The relevant 2-cycle in this case is a disk $C_0$ ending on the topological
D-brane. The real part of
$t^n_i$ measures the size of an $n$-fold cover of this
disk (there is no chemical potential, i.e. $t_0$
is real, since there is associated BPS state of finite mass).
Then \qppl\ follows because
$$
\#(C_0\cap D) = -N.
$$
To see this note that in homology, the class $3N [D_t]$ could equally
well be represented by $-N$ D-branes on the base ${\rm {\bf P}^2}$ and
the latter has intersection number $1$ with $C_0$. The factor of $n$ in
\qppl\ comes about since $t^n$ corresponds to the size of the
${n}$-fold cover of the disk.
\vskip 0.5cm
$ii.\;$ $The$ $general$ $N_i$ $case.$
\vskip 0.5cm
The case $N_1> N_2 > N_3$ is substantially more involved,
and in particular, the large $N$ limit of the amplitudes \intergen\zymP\
is not known.  However, as we will explain in the appendix D, $turning$ $off$
the $U(1)$ factors of the gauge theory, the large $N$ limit can be computed,
and we find a remarkable agreement with the conjecture of \osv .

Let us focus on the leading chiral block of the amplitude.
The large $N$, $M$ limit of the interaction ${V}_{Q R}(M,N)$ (more precisely, the modified version of it to turn off the $U(1)$ charges) is
\eqn\limitnew{\eqalign{
{V}_{\Q \R}\sim \; \beta_M\,\,
&q^{(\vert Q_+ \vert +\vert Q_- \vert) (N-M)\over 2}
q^{(\vert R_+ \vert +\vert R_- \vert) (M-N)\over 2}\cr
&q^{-{(\kappa_{R_+}+\kappa_{R_-}+\kappa_{Q_+}+\kappa_{Q_-}) \over 2}}
{W}_{Q_+ R_+}(q)\;{W}_{Q_- R_-}(q)
}}
where
$$\beta_M = q^{\left(\rho_M^2 +{M\over 24}\right)} M(q^{-1}) \eta^M(q^{-1})\theta^M(q)
$$


In \limitnew\ the $W_{PR}$ is related to the topological vertex amplitude as
$W_{PR}(q) =(-)^{\vert P\vert+\vert R\vert} \hat{C}_{0P^TR}(q)q^{\kappa_R/2}$. It is easy to see that for $N=M$
this agrees with the large $N$ limit of the simpler form of the $V_{\R\Q}$
amplitude in \transf.
It is easy to see that that the leading chiral block of \zymP\ is
\eqn\resultii{
Z_{qYM} \sim Z_{0,0,0}^+\bigl(t\bigr) Z_{0,0,0}^-\bigl(\bar t \bigr)
}
where $Z_{0,0,0}^{+}(t)$ is
$$
Z^+_{0,0,0}={\hat Z}_0
\sum_{ R_+, Q_+, p_+} {W}_{R_+ Q_+}(q) {W}_{Q_+ P_+}(q)
{W}_{P_+ R_+}(q)
e^{-t(\vert R_+ \vert + \vert Q_+ \vert+
\vert P_+)} $$
which is the closed topological string amplitude on $X$.
In particular, this agrees with the amplitude in \chiriii .
In the present context, the Kahler modulus $t$ is given by
$$
t=\half (N_1+N_2+N_3)g_s  -i\theta.
$$
This is exactly as dictated by the attractor mechanism corresponding to the divisor $[D]=(N_1 +N_2 +N_3) [D_t]$!

The higher chiral blocks will naturally be more involved in this
case. Some of the intersection numbers fixing the attractor
positions of ghost branes are ambiguous, and correspondingly, far
more complicated configurations of non-normalizable modes are
expected.



\subsec{Branes on local ${\rm\bf P}^2$}

In this and subsequent section we will discuss the degeneracies of BPS
states that follow from \zymP .  Using the results of \modanul\ and
\modul\ or by direct computation, it is easy to see that $Z_{qYM}$ is
a modular form. Its form however is the simplest in the case
$$
N_1=N_2=N_3=N,
$$
so let us treat this first.
\vskip 0.5cm
$i.\;$ $Degeneracies$ $for$ $N_i=N.$
\vskip 0.5cm
In this case, the form of the partition function written in \zymii\ is
more convenient.  By trading the sum over representations and over the
Weyl-group, as in \modul , for sums over the weight lattices, the
partition function of BPS states is
\eqn\ymiii{
Z_{qYM}(N,\theta_i,g_s) = \beta \sum_{w \in S_N} (-)^{w} \sum_{n_1,n_2,n_3 \in Z^N}
q^{\half \sum_{i=1}^3 {n_i^2 }}\, \, q^{w(n_1)\cdot n_2+n_2\cdot n_3+
n_3\cdot n_1}\, \, e^{i \sum_{i=1}^3 { \theta}_i e(N)\cdot n_i}
}
where $e(N)=(1,\ldots,1)$ and $\beta=\alpha \theta^{3N}(q).$ 
The amplitudes depend on the permutations
$w$ only through their conjugacy classes, consequently we have:
\eqn\ymiv{
Z_{qYM}=\beta \sum_{{\vec K} } d({\vec K})\,\,
 {Z}_{K_1} \times \ldots \times { Z}_{K_r} }
where ${\vec K}$ labels a partition of $N$ into natural numbers
$N=\sum_{a=1}^r K_a,$
and $d({\vec K})$ is the number of elements in the conjugacy class
of $S_N$, the permutation group of $N$ elements, corresponding to having $r$
cycles of length $K_a$, $a=1,\ldots, r$,
and
\eqn\defzk{
{ Z}_{K}(\theta_i,g_s)=(-)^{w_K} \sum_{n_1,n_2,n_3 \in {\bf Z}^K}
q^{{1\over 2}\sum_{i=1}^3 n_i^2 }\, \, q^{w_K(n_1)\cdot n_2+n_2\cdot n_3+
n_3\cdot n_1}\, \, e^{i \sum_{i=1}^3 { \theta}_i e(K)\cdot n_i} }
Here $w_K$ stands for cyclic permutation of K elements. Note that the
form of the partition function \ymiv\ suggests that $Z_{qYM}$
is counting not only BPS bound states, but also contains contribution
from marginally bound states corresponding to splitting of the $U(N)$
to
$$
U(N) \rightarrow U(K_1)\times U(K_2)\times \ldots \times U(K_r)
$$
In each of the sectors, the quadratic form is degenerate.
The contribution of bound states of $N$ branes $Z_N$ diverges as
$$
Z_N(\theta_i,g_s)\sim \sum_{m_1,m_2\in{\rm \bf Z}}
e^{i N m_1 (\theta_1-\theta_3)} e^{i N m_2(\theta_2-\theta_3) }=
\delta\bigl(N(\theta_1-\theta_3)\bigr)
\delta\bigl(N (\theta_2-\theta_3) \bigr)
$$
This is exactly the type of the divergence we found at large $N$ in
the previous subsection.  This divergence should be related to summing
over $D_2$ branes with charges in $H_{2}(D, {\rm \bf Z})^{\perp}$ --
these apparently completely decouple from the rest of the theory.

More precisely, writing $U(N) = U(1)\times SU(N)/{\rm \bf Z}_N$, this
will have a sum over 't Hooft fluxes which are correlated with the
fluxes of the $U(1)$.  Then, $Z_N$ is a sum over sectors of different
$N$-ality,
$$
{ Z}_{N}(\theta_i, g_s)=(-)^{w_N}\sum_{L_{i}=0}^{N-1} \; \sum_{\ell_i \in {\bf Z}+ {{L}_i\over N}} q^{{N\over 2}(\ell_1+\ell_2+\ell_3)^2} e^{i N\sum_i \theta_i \ell_i}
\sum_{m \in {\bf Z}^{3(N-1)} + {\vec \xi}(L_i)}
q^{{1\over 2} m^T {\cal M}_{N} m }
$$
where ${\cal M}_N$ is a non-degenerate
$3(N-1)\times 3(N-1)$ matrix with integer entries  and
$\vec{\xi_i}$ is a shift of the weight lattice corresponding to turning on 't Hooft flux. Explicitly,
$$
\xi^a_i={N-a\over N}L_i,\quad i=1,2,3 \quad a=0,\ldots N-1
$$
where ${\cal M}_N$ is $3(N-1)\times 3(N-1)$ matrix
\eqn\mk{
{\cal M}_{N}=
\pmatrix{M_N & W_{N} & M_N \cr W_{N}^T & M_N & M_N \cr M_N & M_N & M_N }
}
whose entries are
\eqn\mkK{
{M}_{N}=
\pmatrix{\;2 & -1 & \;0 & \ldots &\;0 &\;0 \cr
-1 & \;\;2 & -1& \ldots &\;0 & \;0 \cr
\;0 & -1 & \;2 & \ldots  &\;0&\; 0 \cr
\;. & \;. & \;. &\ldots  &\;.&\;. \cr
\;. & \;. & \;. & \ldots  &\;. &\;. \cr
\;0 & \;0 & \;0 & \ldots  &-1 &\;2 }
}
and
\eqn\mKKK{
{W}_{N}=
\pmatrix{-1 & \;2 &  -1& \ldots & \; 0 &\;0 \cr
        \;0 & -1 & \;2& \ldots & \; 0& \;0 \cr
        \;0 & \;0 & -1 & \ldots &\; . &\;0 \cr
        \;. & \;. & \;. & \ldots & \;. &\;. \cr
        \;. & \;. & \;. & \ldots & \; .&\;. \cr
         -1 & \;0 & \;0 & \ldots& \;0&-1 }
}

We can express $Z_N$ in terms of  $\Theta$-functions
$$
{Z}_N(\theta_i,g_s) =(-)^{w_N} \delta\bigl(N (\theta_1-\theta_3)\bigr)
\delta\bigl(N(\theta_2-\theta_3)\bigr)
\sum_{L_i=0}^{N-1}\;
\Theta_1[a(L_i),b](\tau) \;
\Theta_{3N-3}[{\bf{a}}(L_i),{\bf{b}}]({\hat \tau})
$$
where
$$
\Theta_k[a,b](\tau) = \sum_{n\in {\bf Z}^k}
 \;e^{\pi i \tau (n+a)^2}\; e^{2\pi i n b}
$$
and
$$
\tau = {i g_s\over 2\pi} N, \qquad
{\hat \tau} ={i g_s\over 2\pi} {{\cal M}}_N
$$
$$
a={L_1+L_2+L_3\over N},\quad b= {N\over 2\pi}\theta, \qquad
{\bf{a}}_L = {\vec {\xi}}(L),\quad {\bf{b}} = {\bf 0},
$$
The origin of the divergent factor we found is now clear: from the
gauge theory perspective it simply corresponds to a partition
function of a $U(1) \in U(N)$ gauge theory on a $4-$manifold whose
intersection matrix is degenerate: $\#(C_i\cap C_j)=1$,
$i,j=1,2,3$.  More precisely, to define the intersection form of
the reducible four-cycle $D$, note that $D$ is homologous to the
(punctured) ${\rm \bf P}^2$ in the base, with precisely the
intersection form at hand. The contribution of marginally bound
states with multiple $U(1)$ factors have at first sight a worse
divergences, however these can be regularized by $\zeta$-function
regularization to zero.\foot{For example, $Z_{N-M}(\theta_i,g_s)
Z_{M}(\theta_i,g_s) \sim \delta\bigl(k(\theta_1-\theta_3)\bigr)
\times \sum_{n\in {\rm{\bf Z}}} 1 \times
\delta\bigl(k(\theta_2-\theta_3) \bigr) \times \sum_{n\in {\rm{\bf
Z}}} 1$. where $k$ is the least common divisor of $N,M$. Using
$\zeta(2s) = \sum_{n=1}^{\infty} {1/ n^{2s}},$ where
$\zeta(0)=-{1\over 2}$, we can regularize $\sum_{n\in \rm\bf{Z}} 1
=0.$} This is a physical choice, since in these sectors we expect
the partition function to vanish due to extra fermion zero modes
\VW\Estrings.

To extract the black hole degeneracies we use that the matrix ${\cal M}_{N}$
is non-degenerate and do modular S-transformation using
$$
\Theta[a,b](\tau)=det(\tau)^{-\half} e^{2\pi i ab}
\Theta[b,-a](-\tau^{-1})
$$
This brings $Z_N$ to the form
$$
Z_N(\theta_i,g_s)=\delta\bigl(N(\theta_1-\theta_3)\bigr)
\delta\bigl(N(\theta_2-\theta_3)\bigr)
(-)^{w_N}  \;
\left( {2 \pi \over  N g_s}\right)^{\half}
\;\left({2 \pi \over  g_s}\right)^{3(N-1) \over 2}\;det^{-\half}{\cal M}_N $$
$$
\sum_{L_i=0}^{N-1}
\sum_{\ell \in {\bf Z}}e^{-{2\pi^2 \over Ng_s} (\ell +{N \theta \over 2\pi})^2} e^{-{2 \pi i (L_1+L_2+L_3) \over N}\ell}
\sum_{m \in {\bf Z}^{3(N-1)}} e^{-{2\pi^2 \over g_s} m^T {\cal M}_N^{-1} m}
e^{-2\pi i m \cdot \xi(L_i)}
$$
where ${\cal M}_N$ is the matrix in \mk .
%
%
\vskip 0.5cm
$ii.$ $Degeneracies$ $for$ $N_1 > N_2 > N_3.$
\vskip 0.5cm
When the number of branes is not equal the partition sum
$Z_{qYM}$ is substantially more complicated. By manipulations similar to the ones in appendix B,
$Z^{qYM}$ can be written as:
$$ Z_{qYM}=\alpha \theta^{N_2+2N_3}(q)\sum_{\nu \in S_{N_1}} (-)^{\nu} \sum_{n_1 \in {\bf Z}^{N_1}}
 \sum_{n_2 \in {\bf Z}^{N_2}}
\sum_{n_3 \in {\bf Z}^{N_3}}
q^{\half(n_1^2-n_3^2)} q^{n_2\cdot \nu(n_1)+n_3\cdot n_2+n_3\cdot n_1}
$$
$$ q^{-{\half n_1 \left( \nu^{-1} {\hat P}_{N_1|N_2} \nu\right)n_1}}
 q^{-{\half n_2 {\hat P}_{N_2|N_3} n_2}}
 q^{-{\half n_1 {\hat P}_{N_1|N_3} n_1}} $$
$$q^{-{\nu(n_1)\cdot (\rho_{N_1}-\rho_{N_2})}}
q^{-{n_2 \cdot (\rho_{N_2}-\rho_{N_3}) }}
q^{-{n_1 \cdot (\rho_{N_1}-\rho_{N_3}) }}
e^{i \theta_1 e(N_1)\cdot n_1+
i \theta_2 e(N_2)\cdot n_2+i \theta_3 e(N_3)\cdot n_3} $$
where operator ${\hat P}_{N|M}$ projects $N$-dimensional vector
on its first $M$ components.

For example, consider $N_1=3, N_2=2, N_3=1.$
In this case there are six terms in the sum
$$ \nu_1=\pmatrix{1 & 0 & 0 \cr 0 & 1 & 0 \cr
0 & 0 & 1 \cr},\quad  \nu_2=\pmatrix{0 & 1 & 0 \cr 1 & 0 & 0 \cr
0 & 0 & 1 \cr},\quad  \nu_3=\pmatrix{1 & 0 & 0 \cr 0 & 0 & 1 \cr
0 & 1 & 0 \cr} $$
$$ \nu_4=\pmatrix{0 & 0 & 1 \cr 0 & 1 & 0 \cr
1 & 0 & 0 \cr},\quad  \nu_5=\pmatrix{0 & 1 & 0 \cr 0 & 0 & 1 \cr
1 & 0 & 0 \cr},\quad  \nu_6=\pmatrix{0 & 0 & 1 \cr 1 & 0 & 0 \cr
0 & 1 & 0 \cr} $$
In this simple case $Z_{qYM}$ has the form
$$Z_{qYM}=\alpha \theta^4(q') \left( {  \pi \over g_s} \right)^2
\Bigl(Z_1-Z_2-Z_3-Z_4+Z_5+Z_6\Bigr)
\quad q'=e^{-{\pi^2 \over g_s}}$$
where
$$ Z_i=
\left( { 2 \pi \over g_s} \right)^3 det^{-\half} {\cal M}_{(i)}
\sum_{f \in {\bf Z}^6}
e^{-{2\pi^2 \over g_s} (f+\Lambda_{(i)})^T {\cal M}_{(i)}^{-1}
(f+\Lambda_{(i)}) } $$
where non-degenerate matrices ${\cal M}_{(i)}$ for $i=1,\ldots 6$
are given by
$${\cal M}_{(1)}=\pmatrix{1 & 0 & 0 & 0 & 0 & 0 \cr
0 & -1 & 0 & 1& 0 & 1 \cr
0 & 0 & 0 & 0 & 1 & 0\cr
0 & 1 & 0 & -1 & 0 & 1 \cr
0 & 0 & 1 & 0 & 0 & 0\cr
0 & 1 & 0 & 1 & 0 & -1},\quad
{\cal M}_{(2)}=\pmatrix{1 & 0 & 0 & 0 & 0 & 0 \cr
0 & -1 & 0 & 1& 0 & 1 \cr
0 & 0 & 0 & 0 & 1 & 0\cr
0 & 1 & 0 & 0 & 0 & 0 \cr
0 & 0 & 1 & 0 & -1 & 1\cr
0 & 1 & 0 & 0 & 1 & -1}$$

$${\cal M}_{(3)}=\pmatrix{0 & 0 & 0 & 1 & 0 & 0 \cr
0 & -1 & 0 & 0 & 1 & 1 \cr
0 & 0 & 1 & 0 & 0 & 0\cr
1 & 0 & 0 & 0 & 0 & 0 \cr
0 & 1 & 0 & 0 & -1 & 1\cr
0 & 1 & 0 & 0 & 1 & -1},\quad
{\cal M}_{(4)}=\pmatrix{0 & 0 & 0 & 0 & 1 & 0 \cr
0 & 0 & 0 & 0 & 0 & 1 \cr
0 & 0 & 0 & 1 & 0 & 0\cr
0 & 0 & 1 & 0 & 0 & 0 \cr
1 & 0 & 0 & 0 & -1 & 1\cr
0 & 1 & 0 & 0 & 1 & -1}$$

$${\cal M}_{(5)}=\pmatrix{0 & 0 & 0 & 1 & 0 & 0 \cr
0 & 0 & 0 & 0 & 0 & 1 \cr
0 & 0 & 0 & 0 & 1 & 0\cr
1 & 0 & 0 & 0 & 0 & 0 \cr
0 & 0 & 1 & 0 & -1 & 1\cr
0 & 1 & 0 & 0 & 1 & -1},\quad
{\cal M}_{(6)}=\pmatrix{0 & 0 & 0 & 0 & 1 & 0 \cr
0 & -1 & 0 & 1 & 0 & 1 \cr
0 & 0 & 1 & 0 & 0 & 0\cr
0 & 1 & 0 & 0 & 0 & 0 \cr
1 & 0 & 0 & 0 & -1 & 1\cr
0 & 1 & 0 & 0 & 1 & -1}$$

and vectors $\Lambda_{(i)}$ for $i=1,\ldots, 6$ have components
$$\Lambda_{(1)}=
{1 \over 2\pi}
(\theta_1,\theta_1,
\theta_1,\theta_2,
\theta_2,\theta_3)+{i g_s \over 2\pi} (2,-{3\over 2},-{1\over 2},-{1\over 2},\half,0)
$$
$$\Lambda_{(2)}^a=\Lambda_{(1)}^a,\quad a=1,\ldots,6 $$
$$\Lambda_{(3)}^1={1 \over 2\pi} (\theta_1,
\theta_1,\theta_1,\theta_2,\theta_2,
\theta_3)+{i g_s \over 2\pi} (\half ,-{3\over 2},1,{1\over 2},-{1\over 2},0)
$$
$$\Lambda_{(6)}^a=\Lambda_{(3)}^a,\quad a=1,\ldots,6 $$

$$\Lambda_{(4)}^1={1 \over 2\pi} (\theta_1,
\theta_1,\theta_1,\theta_2,\theta_2,\theta_3)+{i g_s \over 2\pi} (\half ,0,-{1\over 2},\half,-{1\over 2},0)  $$
$$\Lambda_{(5)}^a=\Lambda_{(4)}^a,\quad a=1,\ldots,6 $$

\newsec{Branes and black holes on local
${{\rm {\bf P}}^1\times {\rm{\bf P}}^1}$ }

For our second example, we will take
a noncompact Calabi-Yau threefold $X$ which is a total space of
canonical line bundle $K$ over the base ${\cal B}={\rm \bf P}^1_{B}\times {\rm \bf P}^1_{F}$
$$
X = K\rightarrow {\rm \bf P}^1_{B}\times {\rm \bf P}^1_{F}
$$
where $K=O(-2,-2).$
The linear sigma model whose Higgs branch is $X$
has chiral fields $X_i$, $i=0,\ldots 4$ and two $U(1)$ gauge fields
$U(1)_B$ and $U(1)_F$ under which the chiral fields have charges
$(-2,1,0,1,0)$ and  $(-2,0,1,0,1)$. The corresponding
D-term potentials are
$$
|X_1|^2+|X_3|^2 = 2|X_0|^2 + r_B
$$
$$
|X_2|^2+|X_4|^2 = 2|X_0|^2 + r_F
$$
The $H^2(X,Z)$ is generated by two classes $[D_F]$ and $[D_B]$.
Correspondingly, there are two complexified Kahler moduli $t_B$ and $t_F$,
$t_B = r_B-i\theta_B$ and $t_F = r_F-i\theta_F$.
There are $4$ ample divisors invariant under the $T^3$ torus actions
corresponding to setting
$$
D_i : X_i=0, \quad i=1,2,3,4
$$
We have that $[D_1] = [D_3] = [D_B]$ and  $[D_2] = [D_4] = [D_F]$.
We take $N_1$ and $N_2$ D4 branes on $D_1$ and $D_3$, and $M_1$
and $M_2$ D4 branes on $D_2$ and $D_4$ respectively, corresponding
to a divisor
$$
D = N_1 D_1+ M_1 D_2 + N_2 D_3 + M_2 D_4
$$
Since the topology of each $D_i$ is ${O}(-2)\rightarrow {\rm\bf P}^1$
we will get four copies of qYM theory of ${\rm P^1}$ with ranks $N_{1,2}$ and
$M_{1,2}$. In addition, from the matter at intersection we get $4$ sets of
insertions of observables \init .
All together, and assuming $N_{1,2} \geq M_{1,2}$, we have
\vskip 0.07in
\eqn\ymnew{\eqalign{
Z_{qYM}=\gamma \sum_{\R_1,\R_2,\Q_1,\Q_2} &{V}_{\Q_1 \R_1}
{V}_{\Q_2 \R_2} {V}_{\R_1 \Q_2}
{V}_{\R_2 \Q_1}
\; q^{\sum_{i=1}^2 C_2(\R_i) + C_2(\Q_i)} \cr
&e^{i{\theta}_{B,1} C_1(\R_1)+i{\theta}_{B,2} C_1(\R_2)}
e^{i{\theta}_{F,1} C_1(\Q_1)+ i{\theta}_{F,2} C_1(\Q_2)}.}
}
Above, $\R_1,\R_2$ are representationss of $U(N_1)$ and $U(N_2)$ and  $\Q_1,\Q_2$ are representations of $U(M_1)$ and $U(M_2)$, respectively.

\bigskip
\centerline{\epsfxsize 3.0truein\epsfbox{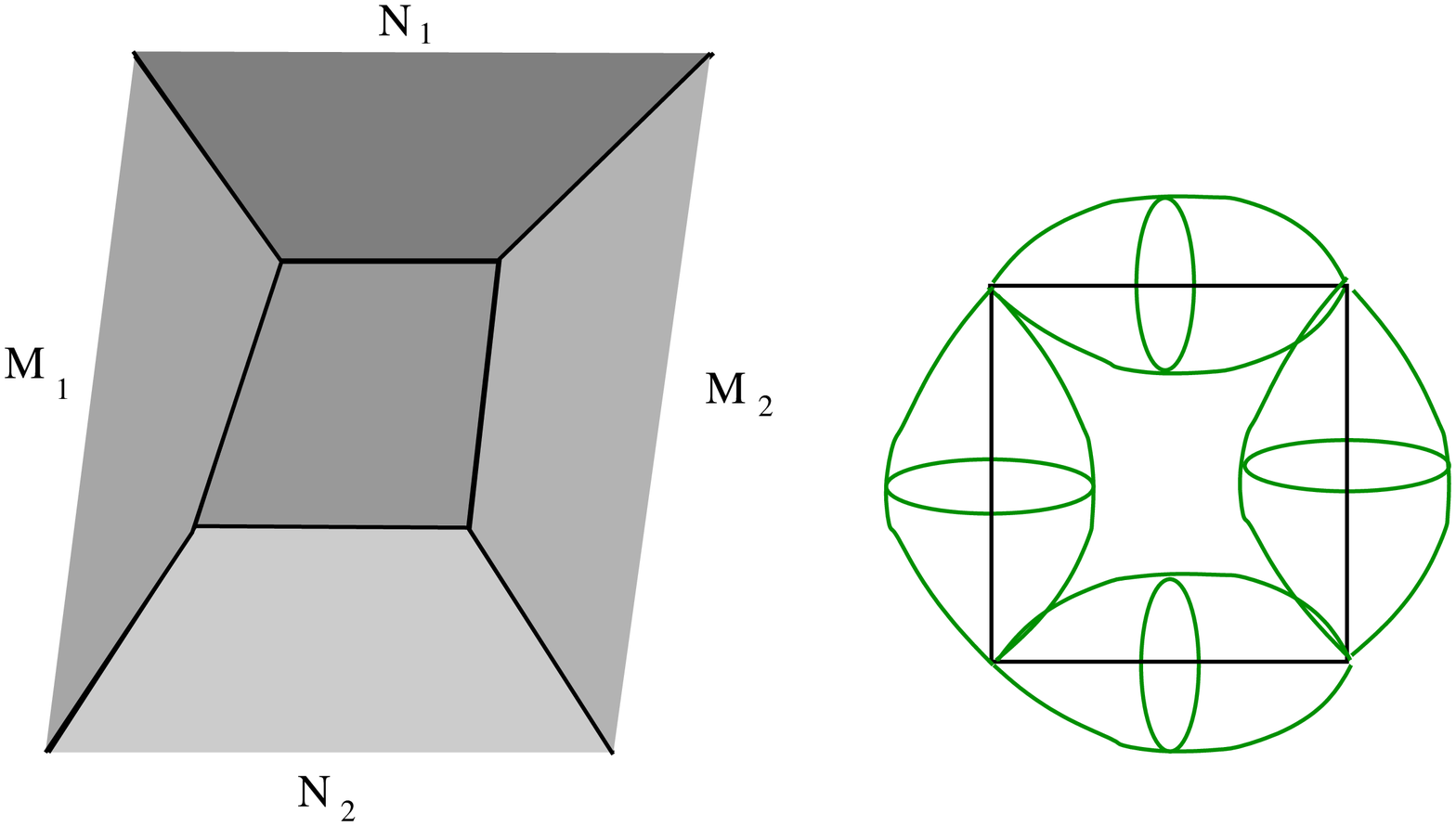}}
\leftskip 2pc
\rightskip 2pc \noindent{\ninepoint\sl
\baselineskip=2pt {\bf Fig. 4.}
{{The base of the local ${\rm \bf  P}^1 \times {\rm \bf P}^1$.
The numbers of D4 branes wrapping the torus invariant non-compact 4-cycles are specified. This corresponds to qYM theory on the neclace of 4 ${\rm\bf P}^1$'s
with ranks $M_1$, $N_1$, $M_2$, and $N_2$.
 }}}
\bigskip

In principle, because $dim(H^2(D,{\rm\bf Z}))=4$,
there 4 different chemical potentials
that we can turn on for the D2 branes, corresponding to
$\theta_{B,i}$, $\theta_{F,i}$. In $X$ however, there are only two independent classes, $dim(H^2(D,{\rm\bf Z}))=2$, in particular
$$
[{\rm\bf P}^1_{B,1}]-[{\rm\bf P}^1_{B,2}]=0,\qquad
[{\rm\bf P}^1_{F,1}]-[{\rm\bf P}^1_{F,2}]=0
$$
We should turn off the chemical potentials for those states that can decay when the YM theory is embedded in string theory, by putting
\eqn\th{
\theta_{B,1} = \theta_{B,2}, \qquad
\theta_{F,1} = \theta_{F,2}.
}
For the most part, we will keep the chemical potentials arbitrary, imposing
\th\ at the end. The prefactor $\gamma $ is
$$ \gamma=q^{-\bigl(2\rho_{M_1}^2 +{M_1 \over 12}\bigr) }
q^{-\bigl(2\rho_{M_2}^2 +{M_2 \over 12}\bigr) } q^{-{1\over 96}
\Bigl( (N_1+N_2)^3+(M_1+M_2)^3 -3(N_1+N_2)^2(M_1+M_2)
-3(M_1+M_2)^2(N_1+N_2) \Bigr)} $$
$$ \times \,\, e^{\theta_B \theta_F (N_1+N_2+M_1+M_2) \over 4g_s}$$
\vskip 0.07in

In the next subsections we will first take the large $N$ limit of the
qYM partition function, and then consider the modular properties of the exact amplitude to compute the degeneracies of the BPS bound states.

\subsec{Black holes on local ${{\rm {\bf P}}^1\times {\rm{\bf P}}^1}$.}

We will now take the large $N$ limit of $Z_{qYM}$ in \ymnew\
and show that this is related to the topological string on $X$
in accordance with the \osv\ conjecture.
\vskip 0.5cm
$i.\;$ $The$ $N_1=N_2=N=M_1=M_2$ $case$.
\vskip 0.5cm
In this case, we can use the simpler form of the vertex amplitude in \intergen\
to write the q-deformed Yang-Mills partition function as:
\eqn\zym{\eqalign{
Z_{qYM}= {\gamma'}\,\sum_{\R_{1,2},\Q_{1,2} \in U(N)}
& S_{\R_1 {\bar\Q_1}}(g_s,N)\, S_{\Q_1{\bar\R_2}}(g_s,N)\,
S_{\R_2 {\bar \Q_2}}(g_s,N)\, S_{\Q_2 {\bar \R_1}}(g_s,N)\cr
&\times e^{i\sum_{i} {\theta}_{B,i} C_1(\R_i)+i{\theta}_{F,i} C_1(\Q_i)}.
}}
where ${\gamma'} =\gamma \theta^{4N}(q)$. Using the large $N$ expansion for S-matrix \smatr\
and parametrizing the $U(1)$ charges $\ell_{R_i}$ of the representations ${\R}_i$ as follows
\eqn\param{
2 \ell_B= \ell_{R_1}+\ell_{R_2},\quad 2\ell_F=\ell_{Q_1}+\ell_{Q_2},\quad
2 n_B= \ell_{R_1}-\ell_{R_2},\quad 2n_F=\ell_{Q_1}-\ell_{Q_2},\quad
}
we find that the sum over $n_{B,F}$ gives delta functions
$$ Z_{qYM}(N,g_s,\theta_{B,i},\theta_{F,i}) \sim \delta \bigl(N (\theta_{B,1}-\theta_{B,2}) \bigr) \;\delta \bigl(N (\theta_{F,1}-\theta_{F,2}) \bigr)\;
Z_{qYM}^{finite}(N,g_s,\theta_B,\theta_F)$$
where
\eqn\partii{\eqalign{
Z_{qYM}^{finite}\sim \sum_{m_B,m_F \in{\rm\bf Z}}
\sum_{P_1,\ldots,P_4}(-)^{\sum_{i=1}^4\vert P_i \vert}
& Z^+_{P_1,\ldots,P_4}\bigl(t_B+m_B g_s,t_F+m_F g_s\bigr)\cr
&Z^+_{P_1^T,\ldots,P_4^T}\bigl({\bar t_B}-m_B g_s,{\bar t_F}-m_F
g_s\bigr)}}
In \partii\
the chiral block  $Z^+_{P_1,\ldots,P_4}(t_B,t_F)$ is given by
\eqn\chirii{\eqalign{
 Z^+_{P_1,\ldots,P_4}(t_B,t_F)=&{\hat Z}_0(g_s,t_B,t_F)
e^{-t_0 \sum_{i=1}^4 \vert P_{i} \vert}
\sum_{R_1,R_2,Q_1,Q_2}
 e^{- t_B (\vert R_1\vert + \vert R_2\vert) }
e^{-t_F (\vert Q_1\vert + \vert Q_2\vert) }\cr
&\times q^{\half \sum_{i=1,2} \kappa_{R_i}+ \kappa_{Q_i}}
 {\hat C}_{Q_1^T R_1 P_1}(q)\, {\hat C}_{R_2^T Q_1 P_2}(q)\,
{\hat C}_{Q_2^T R_2 P_3}(q)\, {\hat C}_{R_1^T Q_2 P_4}(q)
}}
where Kahler moduli are
$$
t_B=g_sN-i\theta_B,\qquad t_F=g_sN-i\theta_F.
$$
The leading chiral block $Z^{+}_{0,\ldots,0}$ is the closed
topological string amplitude on $X.$ The Kahler moduli of the base
${\rm\bf P}^1_B$ and the fiber ${\rm\bf P}^1_F$ are exactly the right
values fixed by the attractor mechanism: since the divisor $D$ that
the D4 brane wraps is in the class $[D]= 2N [D_F]+2N [D_B].$ As we
discussed in the previous section in detail, the other chiral blocks
\chirii\ correspond to having torus invariant non-normalizable modes
excited along the four non-compact toric legs in the normal directions
to the base ${\cal B}$. Moreover the associated Kahler parameters
should also be fixed by the attractor mechanism -- as discussed in the
previous section, we can think of these as the open string moduli
corresponding to the ghost branes. The open string moduli are
complexified sizes of holomorphic disks ending on the ghost branes and
these can be computed using the Kahler form on $X$. Since the net D4
brane charge is the same as that of $-N$ branes wrapping the base, and
the intersection number of the disks $C_0$ ending on the topological
D-branes with the base is $\#(C_0\cap {\cal B})=1$, so the size of all
the disks ending on the branes should be $t_0=-{1\over 2}Ng_s$, which
is in accord with \chirii .  The prefactor in \chirii\ is $$ {\hat
Z}_0(g_s,t_B,t_F)=e^{{1 \over 24 g_s^2}
\bigl( t_F^3+t_B^3 -3t_F^2 t_B-3t_B^2t_F \bigr)}\,\,
M^4(q^{-1}) \eta^{t_B+t_F\over g_s}(q^{-1}) \theta^{t_B+t_F\over g_s}(q)
$$
As discussed before, the eta and theta function pieces contribute
only to the genus zero amplitude, and to the non-perturbative
terms.

\vskip 0.5 cm
$ii.\;$ $The$ general $N_{1,2}$, $M_{1,2}$ $case$.
\vskip 0.5 cm

We will assume here $N_i >M_j$, $i,j=1,2$.
Using the large $N$, $M$ limit of ${V}_{{\cal R} {\cal Q} }(N,M)$
with $U(1)$ charges turned off (see Appendix D) we find
that the leading chiral block of the YM partition function is
$$
Z_{qYM}\sim
Z^+_{0,\ldots,0}\bigl(t_B,t_F\bigr)
Z^-_{0,\ldots,0}\bigl({\overline t_B}, {\overline t_F}
\bigr)
$$
where $Z^+_{0,\ldots,0}\bigl(t_B,t_F\bigr)$ is precisely the topological closed
string partition function on local ${\rm\bf P}^1\times {\rm\bf P}^1$ \AKMV\ :
$$
Z^+_{0,\ldots,0}={\hat Z}_0
\sum_{ Q_1^+, Q_2^+,R_1^+,R_2^+} {W}_{Q_1^+ R_1^+}(q) {W}_{Q_1^+ R_2^+}(q)
{W}_{Q_2^+ R_1^+}(q) {W}_{Q_2^+ R_2^+}(q)
e^{-t_F(\vert Q_1^+ \vert + \vert Q_2^+ \vert)}
 e^{-t_B(\vert R_1^+ \vert + \vert R_2^+ \vert)} $$
It is easy to see that this agrees with the amplitude given in \chirii .
Moreover, the Kahler parameters are exactly as predicted by the attractor mechanism corresponding to having branes on a divisor class
$$[D] =(N_1+N_2)[D_B]+(M_1+M_2)[D_F].$$
Namely,
$$
t_B=\half (M_1+M_2) g_s-i\theta_B, \qquad t_F=\half (N_1+N_2) g_s-i\theta_F.
$$
Note that the normal bundle to each of the divisor $D_i$ is trivial,
so the size of the corresponding ${\rm \bf P}^1$ in $D_i =
O(-2)\rightarrow {\rm \bf P}^1$ is independent of the number of branes
on $D_i$, but it does depend on the number of branes on the adjacent faces
which have intersection number $1$ with the ${\rm \bf P}^1$.

It would be interesting to study the structure of the higher chiral blocks.
In this case we expect the story to be more complicated, in particular
because some of the intersection numbers that compute the attractor values of the brane moduli are now ambiguous.

\subsec{Branes on local ${\rm\bf P}^1 \times {\rm\bf P}^1$}

We will content ourselves with considering $N_{1,2}=M_{1,2} =N$
case, the more general case working in similar ways to the local
${\rm\bf P}^2$ case. The partition function \zym\
may be written as

\eqn\ymviii{
Z_{qYM}(N,\theta_i,g_s) = {\gamma'} \sum_{w \in S_N} (-)^{w}
\sum_{n_1,\ldots,n_4 \in Z^N}
q^{w(n_1)\cdot n_2+n_2\cdot n_3+n_3\cdot n_4 +n_4\cdot n_1}
\, \, e^{i \sum_{i=1}^4 { \theta}_i e(N)\cdot n_i} }
where  $e(N)=(1,\ldots,1)$.
As before in the case of local ${\rm\bf P}^2,$ the
bound states of $N$ D4-branes are effectively counted by the
$Z_N$ term, i.e. the term with $w=w_N$.
Like in that case, $Z_N$ is again a sum over sectors of different
$N$-ality,
$$
{ Z}_{N}(\theta_i, g_s)={\gamma'}\, (-)^{w_N}\sum_{L_1,\ldots,L_4=0}^{N-1} \;
\sum_{\ell_i \in {\bf Z}+ {{L}_i\over N}}
q^{N(\ell_1+\ell_3)(\ell_2+\ell_4)} e^{i N\sum_{i=1}^4 \theta_i \ell_i}
\sum_{m \in {\bf Z}^{4(N-1)} + {\vec \xi}(L_i)}
q^{{1\over 2} m^T {\cal M} m }
$$
where ${\cal M}$ is a non-degenerate
$4(N-1)\times 4(N-1)$ matrix with integer entries  and
$\vec{\xi_i}$ is a shift of the weight lattice corresponding to turning on 't Hooft flux.

More explicitly,
$$ \xi^a_i={N-a\over N}L_i,\quad i=1,\ldots, 4
\quad a=0,\ldots N-1 $$
${\cal M}$ is
$4(N-1)\times 4(N-1)$ matrix
\eqn\mkii{
{\cal M}=
\pmatrix{0 & W_{N} & 0& M_N \cr
W_{N}^T & 0&  M_N & 0 \cr
0& M_N & 0& M_N \cr
M_N & 0 & M_N &0}
}
\vskip 0.3cm
whose entries are $(N-1)\times (N-1)$ matrices

\eqn\mkKii{
{M}_{N}=
\pmatrix{\;2 & -1 & \;0 & \;0& \ldots &\;0 &\;0 \cr
-1 & \;\;2 & -1& \;0&\ldots &\;0 & \;0 \cr
\;0 & -1 & \;2 & -1&\ldots  &\;0&\; 0 \cr
\;. & \;. & \;. & \;.&\ldots  &\;.&\;. \cr
\;. & \;. & \;. & \;.&\ldots  &\;. &\;. \cr
\;0 & \;0 & \;0 & \;0&\ldots  &-1 &\;2 }
}
and
\eqn\mKKKii{
{W}_{N}=
\pmatrix{-1 & \;2 &  -1& \;0&\ldots & \; 0 &\;0 \cr
        \;0 & -1 & \;2& -1 &\ldots & \; 0& \;0 \cr
        \;0 & \;0 & -1 & \;2&\ldots &\; . &\;0 \cr
        \;. & \;. & \;. & \;.&\ldots & \;. &\;. \cr
        \;. & \;. & \;. & \;.& \ldots & \; .&\;. \cr
        \;0 & \;0 & \; 0 & \; 0&\ldots &-1 &\;2 \cr
         -1 & \;0 & \;0 & \; 0&\ldots& \;0&-1 }
}
We can express $Z_N$ in terms of  $\Theta$-functions
$$\eqalign{
{Z}_N(\theta_i,g_s) =& {\gamma'}\, (-)^{w_N}
\delta\bigl(N (\theta_{B,1}-\theta_{B,2})\bigr)
\delta\bigl(N(\theta_{F,1}-\theta_{F,2})\bigr)\cr
&\sum_{L_1,\ldots,L_4 =0}^{N-1}\; \;
\Theta_2[a(L_i),b](\tau) \; \;
\Theta_{4N-4}[{\bf{a}}(L_i),{\bf{b}}]({\hat \tau})
}
$$
where
$$
\Theta_k[a,b](\tau) = \sum_{n\in {\bf Z}^k}
 \;e^{\pi i \tau (n+a)^2}\; e^{2\pi i n b}
$$
and
$$
\tau = {i g_s\over 2\pi} N\pmatrix{ 0 & 1 \cr 1 & 0}, \qquad
{\hat \tau} ={i g_s\over 2\pi} {{\cal M}}
$$
and
$$
a=\bigl({L_1+L_3\over N},{L_2+L_4\over N}\bigr),
\quad b= \bigl( {N\over 2\pi}\theta_B,{N\over 2\pi}\theta_F\bigr) \qquad
{\bf{a}}_L = {\vec {\xi}}(L),\quad {\bf{b}} = {\bf 0},
$$
To extract black hole degeneracies  we  use that matrix ${\cal M}$
is non-degenerate and do modular S-transformation using
$$
\Theta[a,b](\tau)=det(\tau)^{-\half} e^{2\pi i ab}
\Theta[b,-a](-\tau^{-1})
$$
After modular S-transformation $Z_N$ is brought to the form
$$\eqalign{
Z_N(\theta,g_s)=& {\gamma'}\, \delta\bigl(N (\theta_{B,1}-\theta_{B,2})\bigr)
\delta\bigl(N(\theta_{F,1}-\theta_{F,2})\bigr)
(-)^{w_N}  \;
\left( {2 \pi \over  N g_s}\right)
\;\left({2 \pi \over  g_s}\right)^{4(N-1) \over 2}\;det^{-\half}{\cal M}\cr
& \sum_{L_1,\ldots,L_4 =0}^{N-1}
\sum_{\ell,\ell' \in {\bf Z}}e^{-{\pi^2 \over Ng_s}
(\ell +{N \theta_B \over 2\pi})
(\ell' +{N \theta_F\over 2\pi})  } e^{-{2 \pi i (L_1+L_3) \over N}\ell}
e^{-{2 \pi i (L_2+L_4) \over N}\ell'}\cr
&\qquad \quad \;\sum_{m \in {\bf Z}^{4(N-1)}}
e^{-{2\pi^2 \over g_s} m^T {\cal M}^{-1} m}
e^{-2\pi i m \cdot \xi(L_i)}
}
$$

\newsec{Branes and black holes on $A_k$ ALE space}

Consider the local toric Calabi Yau $X$ which is $A_k$ ALE space times
${\rm \bf C}$.  This can be thought of as the limit of the usual ALE
fibration over ${\rm \bf P}^1$ as the size of the base ${\rm \bf P}^1$
goes to $\infty$.  In this section we will consider black holes
obtained by wrapping $N$ D4 branes on the ALE space.

\bigskip
\centerline{\epsfxsize 4.0truein\epsfbox{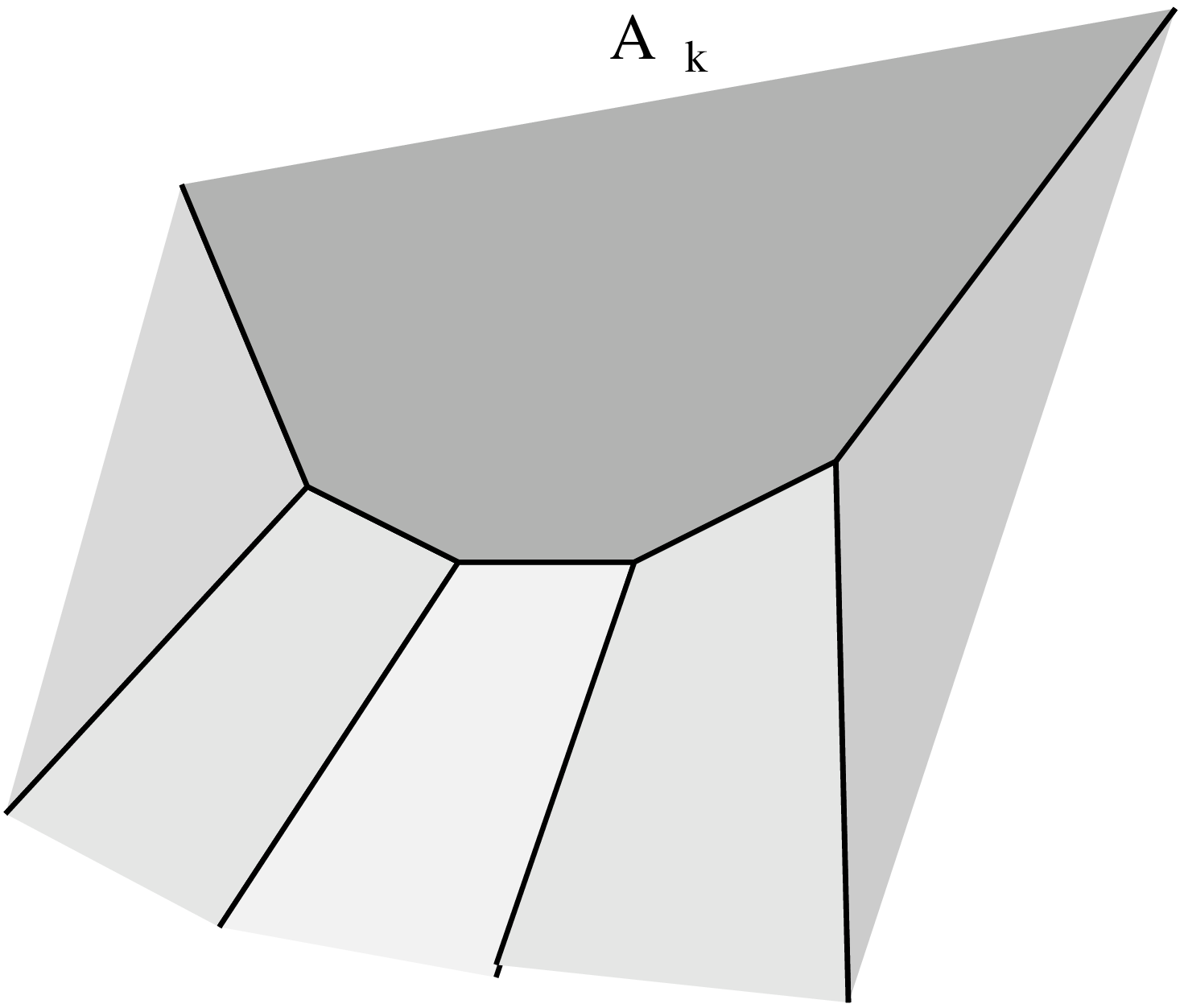}}
\leftskip 2pc
\rightskip 2pc \noindent{\ninepoint\sl
\baselineskip=2pt {\bf Fig. 5.}
{{$N$ D4-branes are wrapped on $A_{k}$ type ALE space in
$A_k\times {\rm \bf C},$ for $k=3$. The D-brane partition function is computed by $U(N)$ qYM theory on a chain of 3 ${\rm\bf P}^1$'s.}}}
\bigskip

This example will have a somewhat different flavor than the
previous two, so we will discuss the D4 brane gauge theory on a
bit more detail.  On the one hand, the theory on the D4 brane is a
topological $U(N)$ Yang-Mills theory on $A_k$ ALE space which has
been studied previously \nakajima\VW . On the other hand, the
$A_k$ ALE space has $T^2$ torus symmetries, so we should be able
to obtain the corresponding partition function by an appropriate
computation in the two dimensional qYM theory.  We will start with
the second perspective, and make contact with \nakajima\VW\ later.

As in \AOSV\ and in section 3, our strategy will be to cut the four
manifold into pieces where the theory is simple to solve, and then
glue the pieces back together.  The $A_k$ type ALE space can be
obtained by gluing together $k+1$ copies of ${\rm \bf C}^2$.
Correspondingly, we should be able to obtain YM amplitudes on the ALE
space by sewing together amplitudes on ${\rm \bf C^2}$.  Moreover,
since the ${\rm \bf C^2}$ and the ALE space have $T^2$ isometries, the
4d gauge theory computations should localize to fixed points of these
isometries, and these are bundles with second Chern class localized at
the vertices, and first Chern class along the edges.

Viewed as a manifold fibered by 2-tori $T^2$, ${\rm \bf C}^2$ has
contains two disks, say $C_{base}$ and $C_{fiber}$ that are fixed
by torus action (see figure 2 by way of example). Viewed as a line
bundle over a disk $C_{base}$ as a base, the $U(1)$ isometry of
the fiber allows us to do some gauge theory computations in the
qYM theory on $C_{base}$. In particular, if the bundle is flat the
qYM partition function on a disk \disk\ with holonomy $U = \exp{(i
\oint A)}$ fixed on the boundary of the $C_{base}$ fixed and no
insertions is\foot{More precisely, as  we explained in section 3,
the coordinate $U$ is ambiguous since the choice of cycle which
remains finite is ambiguous. This ambiguity relates to the choice
of the normal bundle to the disk, and the present choice
corresponds to picking this bundle to be trivial, which is
implicit in the amplitude.}
$$
Z(C) (U)= \sum_{\R} e^{i \theta C_1(\R)}
S_{0\R}(N,g_s) Tr_{\R}U.
$$
What is the four
dimensional interpretation of
this? The sum over $\R$ in the above corresponds
to summing over the four dimensional $U(N)$ gauge fields with
\eqn\shift{
\int_{fiber} F_a = \R_a \, g_s, \qquad a=1,\ldots N,
}
where ${\R}_a$ are the lengths of the rows in the Young tableau of
$\R$.\foot{To be more precise, $\R_a$ in \shift\ is shifted by
${1\over 2}(N+1)-a.$} This is because on the one hand \eqn\oper{
S_{0\R}(N,g_s) = \langle \;Tr_{\R}\, e^{i\oint A}\; \rangle. } and
on the other $\oint A_a=\int_{base} F_a$ is conjugate to $\Phi_a =
\int_{fiber} F_a$, so inserting \oper\ shifts $F$ as in \shift .
The unusual normalization of $F$ has to do with the fact that qYM
directly computes the magnetic, rather than the electric partition
function: In gluing two disks to get an ${\rm\bf P}^1$ we sum over
all $\R$'s labelling the bundles of the S-dual theory over the
${\rm\bf P}^1$.

If we are to use 2d qYM theory to compute the
${\cal N}=4$ partition function on ALE space, we must understand what
in the 2d language is computing the partition function on
${\rm \bf C}^2$ with
\eqn\flux{
\int_{fiber} F_a = \R_a \, g_s, \qquad
\int_{base} F_a = \Q_a \, g_s, \qquad a=1,\ldots N,
}
since clearly, what we call the ``base'' here versus the ``fiber'' is a matter of convention.
Using once more the fact that
$\Phi$ and $\oint A$
are conjugate,  turning on
$\int_{base} F_a = \Q_a g_s$ corresponds to inserting
$Tr_{\Q}\, e^{-i \Phi}$ at the point on $C_{base}$ where it intersects $C_{fiber}$. Thus, turning on \flux\ corresponds to computing
$\langle \,
Tr_{Q} e^{-i\Phi}\; Tr_{\R}\, e^{ i\oint A}\, \rangle.
$
This is an amplitude we already know:
\eqn\opertwo{
S_{{Q}\R}(N,g_s) = \langle\,
Tr_{Q} e^{- i\Phi} \, Tr_R e^{ i\oint A}\, \rangle.
}
Alternatively, the amplitude on ${\rm\bf C}^2$ with arbitrary boundary
conditions \flux\ on the base and on the fiber is
\eqn\vv{
\sum_{\R,\Q} \; S_{\R{ \Q}}(N, g_s) \;Tr_{\R} U \;Tr_{\Q}V
}
We then glue the pieces together using the usual local rules.  The
only thing we have to remember is that the normal bundle to each
${\rm\bf P}^1$ is ${O}(-2)$, and that at the ``ends'' we should
turn the fields off.  In computing \opertwo\ we used the
coordinates in which ${\rm\bf C}^2$ is a trivial fibration over both
$C_{fiber}$ and $C_{base}$, and therefore to get the first Chern
class of the normal bundle to come out to be $-2$, we must along each
of them insert annuli with $O(-2)$ bundle over them.  This gives:
\eqn\ALE{ Z = \sum_{{\cal R}_1 \dots
{\cal R}_k} S_{0 {\cal R}_{(1)}} S_{{\cal R}_{(1)} {\cal R}_{(2)}}
\dots S_{{\cal R}_{(k)} 0} \;q^{\sum C_2 ({\cal R}_{(j)})} \;
e^{ i \sum\theta_j |{\cal R}_{(j)}|},}
There is one independent $\theta$ angle for each ${\rm \bf P}^1$ corresponding to the fact that they are all independent in homology. These $\theta$ angles will get related to chemical potentials for the D2 branes wrapping the corresponding 2-cycles.

\subsec{Modularity}

The S-duality of ${\cal N} = 4$ Yang Mills acts on our partition
function as $g_s \rightarrow \frac{4\pi^2}{g_s}$. By performing this
modular transformation we will be able to read off the degeneracies of
the BPS bound states contributing to the entropy. First, using the
definition of the Chern Simons S-matrix, we find that
\eqn\ALE{
Z = \sum_{\omega \in {\cal W}}(-1)^{\omega} \sum_{n_1,\ldots n_k\in Z^{N}} q^{n_1^2 + \ldots+ n_k^2 - n_1 n_2 - \ldots -n_{k-1} n_k} \; e^{i \theta_1 |n_1|+\ldots \theta_k |n_k|} q^{\rho n_1 + n_k \omega(\rho) }
}
Note the appearance of the intersection matrix of $A_k$ ALE space.
The
fact that the Cartan matrix appears gives the $k$ vectors $U(N)$ weight vectors $n_{i}^{a}$ $i=1,\ldots k$, $a=1,\ldots N$ an alternative interpretation as $N$ $SU(k)$ $root$ vectors:
$$
Z =  \sum_{\omega \in {\cal W}}(-1)^{\omega}
\prod_{a=1}^{N} \sum_{n_a \in \Lambda_{SU(k)}^{Root}} q^{{1\over 2} n_a n_a} \; e^{i\theta n_a} q^{ (\rho+\omega(\rho))_a n_a}
$$
where $\theta$ is a $k$-dimensional vector with entries $\theta_i$. From the above, it is clear that $Z$ is a product of $N$ $SU(k)$ characters at level one. Recall that the level one characters are
$$
\chi^{(1)}_{\lambda}(\tau,u) = {\theta^{(1)}_{\lambda}(\tau,u) \over \eta^k(\tau)}
$$
where
$$
\theta^{(1)}_{\lambda}(\tau,u) = \sum_{n \in  \Lambda_{SU(k)}^{Root}}
e^{\pi i \tau (n + \lambda)^2 + 2 \pi i (n+\lambda) u}
$$
To be concrete, our amplitude is given as follows:
$$
Z = \eta(q)^{Nk}\,\, \sum_{\omega \in {\cal W}}(-1)^{\omega}
\prod_{a=1}^{N}
\chi_0^{(1)}(\tau, u^{a}(\theta,\omega))
$$
Here,
$$
\tau = {i g_s \over 2 \pi}, \quad \quad
u_i^a(\theta,\omega)= {\theta_i\over 2 \pi} +{i g_s \over 2 \pi}(\rho +\omega(\rho))^a
$$
Modular transformations act on the space of level one characters as:
$$
\theta^{(1)}_{\eta}(-{1\over \tau}, {u\over \tau}) = e^{-{u u \over 2\tau}}\sum_{\omega \in {\cal W}_k}(-1)^{\omega}\sum_{\lambda}
e^{{2 \pi i \over k+1}\omega(\eta+\rho) (\lambda + \rho)}  \theta^{(1)}_{\lambda}(\tau,u),
$$
consequently, the dual partition function also has an expansion in terms of
$N$ level one characters.
The product of $N$ level one characters can be expanded in terms of
sums of level $N$ characters, so this is consistent with the results
of H. Nakajima.
The fact that the partition function is a sum over level $N$ characters,
rather than a single one is natural given that we
impose different boundary conditions at the infinity of
ALE space from \nakajima .

\subsec{The large $N$ limit}

In the 't Hooft large N expansion, using \smatr , we find that the
partition function \ALE\ can be written as follows:
$$
Z_{ALE} = \sum_{P_1, \dots, P_{k+1}} (-)^{\vert P_1\vert + \dots \vert P_{k+1}\vert}
\sum_{m_1, \dots, m_k \in {\bf Z}}$$
$$Z^+ _{P_1, .. , P_{k+1}}(t_1 + m_1 g_s, \dots, t_k + m_k g_s)\,\,
Z^+_{P_1^T, .. , P_{k+1}^T}({\bar t}_1 - m_1 g_s, \dots,
{\bar t}_k- m_k g_s ),
$$
where
$m$'s are related to the $U(1)$ charges of representations $\R_i$ as
${m}_i = 2\ell_i
-\ell_{i-1} - \ell_{i+1}$, for $i = 1, \dots, k$ (where $\ell_0 = \ell_{k+1} = 0$).
The Kahler moduli are
$$
t_j=-\,i\, \theta_j, \qquad  j=1,\ldots,k,
$$
which is what attractor mechanism predicts: Since ALE space has
vanishing first Chern class, the normal bundle of its embedding in
a Calabi-Yau three-fold is trivial, and consequently $\#[D_{A_k}
\cap C]=0$ where $D_{A_k}$ is ($N$ times) the divisor
corresponding to the ALE space and $C$ is any curve class in $X$.

The normalization constant $\alpha_{ALE}$ in \ALE\
was determined by requiring the large $N$ limit factorizes in the appropriate way.
\eqn\ALEnrm {\alpha_{ALE}=
q^{(k+1)(\rho^2 + \frac{N}{24})} e^{\frac{N}{2 g_s}
\theta^T A  \theta},} where $A$
is the inverse of the intersection matrix of ALE.

The chiral block in the chiral(anti-chiral) decomposition of
$Z_{ALE}$ has the form
$$
Z^{+}_{P_1,\ldots,P_{k+1}}(t_1, \dots, t_k) = M(q)^{\frac{k+1}{2}}
e^{-\frac{t_0\, t^T \, A \, t}{2 g_s^2}+\frac{\pi^2 (k+1) t_0}{6 g_s^2}}
e^{-t_0 \sum_{d=1}^{k+1} |P_d|}\times$$ $$ \times \sum_{R_1
\dots R_k} {\hat{C}}_{0 {R_1}^T P_1} q^{\kappa_{R_1}/2} e^{-t_1 |R_1|}
\hat{C}_{R_1 {R_2}^T P_2} q^{\kappa_{R_2}/2} e^{-t_2 \vert R_2 \vert} \dots
{\hat{C}}_{R_k 0 P_{k+1}}.
$$
where
\eqn\diskALE{
t_0 = {1\over 2} Ng_s.
}

We see that the trivial chiral block $Z^+_{0,\ldots,
0}(t_1,\ldots,t_k)$ is exactly the topological string partition
function on ALE, in agreement with the conjecture of \osv .
Moreover, the higher chiral blocks correspond to having $k+1$ sets
of topological ``ghost`` branes in the ${\rm \bf C}$ direction
over the north and the south poles of the ${\rm \bf P}^1$'s.  The
associated moduli, i.e. the size of the holomorphic disks ending
on the topological ghost branes is also fixed by the attractor
mechanism, to be $\#(D_{A_k}\cap C_{disk}) = N$. This is gives
exactly \diskALE\ as the value of the corresponding Kahler moduli
$t_0$, in agreement with the conjecture. As we discussed in
section 4, in the closed string language, these are the
non-normalizable modes in the topological string on $X$. The
classical piece of the topological string amplitude \eqn\class{
\frac{1}{2 g_s^2}\, t_0\, t^T A  t } deserves a comment. Because
$X=A_k \times {\rm\bf C}$, taking only the compact cohomology the
triple intersection numbers would unambiguously vanish. The
non-vanishing triple intersection numbers can be gotten only by a
suitable regularization of the ${\rm\bf C}$ factor. This $was$
already regularized, in terms of the Kahler modulus $t_0$ of the
non-normalizable modes -- which exactly give the measure of the
size of the disk, i.e. ${\rm \bf C}$, making \class\ a natural
answer.\foot{What is less natural is the appearance of the
$inverse$ intersection matrix of ALE. However, one has to remember
that this is a non-compact Calabi-Yau, where intersection numbers
are inherently ambiguous.}

\break
\vskip 1cm
{\bf Acknowledgments}

We are grateful to C. Vafa for many very valuable discussions, and
collaboration on a related project. We also thank F. Denef, J.
McGreevy, M. Marino, A. Neitzke, T. Okuda, and H. Ooguri for
useful discussions. M.A and N. S. are grateful to the KITP
program on the "Mathematical Structures in String Theory", where
part of this work was done. The research of M.A. is supported in
part by a DOI OJI Award, the Alfred P. Sloan Fellowship and the
NSF grant PHY-0457317. 
Research of N.S. and D.J. was supported in part by NSF
grants PHY-0244821 and DMS-0244464.
\vskip 2cm
\appendix{A}{Conventions and useful formulas}

%

The $S$ matrix is given by
$$
S_{\R\Q}(N,g_s)=
\sum_{w \in S_N} (-)^w q^{-w(\R+\rho_N) \cdot (\Q+\rho_N)}
$$
where $q = exp(-g_s)$, and $\rho^a_N={N-2a+1\over 2}$, for $a=1,\ldots,N.$
Note that while the expression for ${S}_{\R\Q}$ looks like that for the
S-matrix of the $U(N)$ WZW model, unlike in WZW case, $g_s$ is not quantized.
Using Weyl denominator formula $Tr_{\R} x = \prod_{i<j}(x_i-x_j) \sum_{w \in S_N} (-)^w x^{w(\R+\rho_N)}$,
the $S$-matrix can also be written in terms of Schur functions
$s_\R(x_1,\ldots,x_N)=Tr_{\R} x $ of $N$ variables.
$$
S_{\R\Q}/S_{00}(g_s,N) = s_\R (q^{-\rho_N-\Q}) s_\Q (q^{-\rho_N}).
$$
Above, $x$ is an $N$ by $N$ matrix with eigenvalues $x_i$, $i=1,\ldots N$, as

The $S$ matrix has following important properties:
$$
S_{{\bar \R} \Q}(N,g_s)=
S_{ \R \Q}(N,-g_s) = S^{-1}_{{\R} \Q}(N,g_s)
$$
The first follows since (up to a sign that is $+1$ if $N$ is odd and $-1$ if $N$ is even), $\bar \Q+\rho_N = -\omega_N (\Q + \rho_N)$ where $\omega_N$ is the permutation that maps $a \rightarrow N-a+1$ for $a=1,\ldots,N.$
The second is easily seen by computing
$$\eqalign{
\sum_{\P} S_{\R \P}(N,-g_s) S_{\P \Q} (N,g_s) &=
 \sum_{w \in S_N} (-)^w \sum_{n \in \bf{Z}^N} q^{w(\rho_N+\R) \cdot n} q^{-n \cdot (\rho_N + \Q)}\cr
&=
\sum_{w \in S_N} (-)^w \delta^{(N)} \left( w(\rho_N+\R)-(\rho_N+\Q) \right) = \delta_{\R \Q}.}
$$
where we absorbed one sum over the Weyl group into the unordered
vector, $n$. Note that $(\rho_N+\R)^a$ and $(\rho_N+\Q)^a$ are
decreasing in $a$, so the delta function can only be satisfied when
$w= 1$.

The large $N$ limit of the $S$ matrix for coupled representations
$\R = R_+\bar{R}_-[\ell_R]$, $\Q = Q_+\bar{Q}_-[\ell_Q]$
is given in \smatr\ in terms of the topological vertex amplitude
$$
{\hat{C}}_{R Q P}(q) = C_{RQP}(q^{-1}),\quad
 C_{RQP}(q)=q^{\kappa_{R}\over 2} s_{P}(q^\rho)
\sum_{\eta} s_{R^t/\eta}(q^{P+\rho}) s_{Q/\eta}
(q^{P^t+\rho})
$$
This has cyclic symmetry ${\hat C}_{PQR} ={\hat C}_{QRP}$,
and using
the properties of the Schur functions under $q \rightarrow q^{-1}$:
$
s_{R}(q^{Q+\rho})= (-1)^{\vert R\vert }s_{R^T}(q^{-Q^T-\rho})
$
also a symmetry under inversion:
$
{\hat{C}}_{R Q P} (q^{-1}) = (-)^{|R|+|Q|+|P|} {\hat{C}}_{R^t Q^t P^t} (q).
$
The leading piece of $S$ in the large $N$ limit is significantly simpler than \smatr . Since
$
{\hat C}_{0R Q}(q) = (-)^{\vert R\vert +\vert Q\vert}
W_{R^TQ}(q)q^{-\half \kappa_Q}
$
we have:
$$\eqalign{
S_{\R\Q}(g_s,N) = &(-)^{\vert R_+\vert +\vert Q_+\vert+\vert
R_-\vert +\vert Q_- \vert} q^{-N \ell_R\ell_Q} q^{-\ell_R (\vert
Q_+\vert -\vert Q_-\vert)} q^{-\ell_Q (\vert R_+\vert - \vert R_-
\vert)}\cr &W_{R_+Q_+}(q) W_{R_-Q_-}(q) q^{-{N\over 2}(|R_+|
+|R_-| +|Q_+| +|Q_-|)}}
$$
where
$$
W_{RQ}(q)=s_R (q^{\rho+Q}) s_Q (q^\rho)
$$
where $\rho=-a+\half,$ for $a=1,\ldots,\infty.$

\appendix{B}{Quantum Yang-Mills amplitudes with observable insertions}

Consider the $U(N)$ q-deformed YM path integral on the cap.
As shown in \AOSV\ this is given by
$$
Z_{{\rm qYM}}(C)(U) = \sum_{\R} S_{0\R} Tr_{\R} U.
$$
The Fourier transform to the $\Phi$ basis is given by the following path
integral over the boundary of the disk,
$$
Z_{{\rm qYM}}(C)(U) = \int
d_H \Phi \;e^{{1\over g_s} Tr \Phi \oint A}\; Z_{{\rm qYM}}(C)(\Phi).
$$
Since the qYM path integral localizes to configurations where $\Phi$ is
covariantly constant,so in particular $\Phi$ and $A$ commute,
integrating over the angles gives\foot{There was an error in \AOSV\
where the denominator ${1/ \Delta_H(u)}$ was dropped. In that case
this only affected the definition of the wave function (whether one
absorbs the determinant $\Delta_H(\phi)$ into the wave function of
$\phi$ or not), but here we need the correct expresion. This
normalization follows from \MM\
where the matrix model for a pair of
commuting matrices with haar measure was first discussed.}
$$
Z_{{\rm qYM}}(C)({\vec u}) = \int \; \prod_i d\phi_i\;
{
\Delta_H(\phi) \over \Delta_H(u)}
\;e^{{1\over g_s}\sum_i{{\vec{\phi}\cdot{\vec u}}}}\;
Z_{{\rm qYM}}(C)({\vec \phi}),
$$
where we defined a hermitian matrix $u$ by
$ U = e^{iu},$
and
$$
\Delta_H(\phi)= \prod_{1\leq i<j\leq N}2sin[(\phi_i-\phi_j)/2]=\prod_{\alpha >0}2\sin({\vec\alpha} \cdot {\vec \phi}).
$$
comes from the hermitian matrix measure over ${\vec \phi}$ by
adding images under ${\vec \phi} \rightarrow {\vec \phi} + 2\pi {\vec n}$,
to take into account the periodicity of $\Phi$.

Now, in the $\Phi$ basis, the path integral on the disk with insertion of
$Tr_{\Q} e^{i\Phi}$ is simply given by:
$$
Z(C, Tr_{\Q} e^{i\Phi})(\Phi) = Tr_{\Q} e^{i\Phi}
$$
since $\Phi$ is a multiplication operator in this basis.
Transforming this to $U$-basis, we use
$$
Tr_\Q e^{i\Phi} := \chi_\Q(\vec{\phi})=
{\sum_{\omega \in S_N} (-1)^{\omega}
e^{i\omega({\vec \Q}+{\vec \rho})\cdot {\vec \phi}}
\over {\sum_{\omega \in S_N} (-1)^{\omega}
e^{i\omega({\vec \rho})\cdot {\vec \phi}}}},
$$
where $S_N$ is the Weyl group and $\vec{\rho}$ is the Weyl vector.
We also use the Weyl denominator formula
$$
\prod_{\alpha>0} \sin({\vec\alpha}\cdot{\vec \phi})=
{\sum_{\omega \in S_N} (-1)^{\omega} e^{i\omega({\vec \rho})\cdot {\vec \phi}}}.
$$
Plugging this into the integral, and performing a sum over the weight lattice
we get
$$
Z(C,Tr_\Q e^{i\Phi})(U) = {1\over \Delta_H(u)}
\sum_{\omega \in W} (-1)^{\omega}
\delta({\vec u} + i g_s{\omega(\vec{\rho}+\vec{\Q})})
$$
We can extract the coefficient of this in front of $Tr_{\R} U$
by computing an integral
$$
\int dU Z(C,Tr_\Q e^{i\Phi})(U) Tr_{\R} U^{-1}
$$
which easily gives
$$
Z(C,Tr_\Q e^{i\Phi})(U) = \sum_{\R} S_{\R{{\bar \Q}}}(g_s,N) Tr_{\R}U.
$$
where
$$
S_{\R {\bar \Q}}(g_s,N) = \sum_{\omega} q^{\omega(\Q+\rho) \cdot (\R +\rho)}
$$
in terms of $q=e^{-g_s}$.

Another expectation value we need is of
$$
Z(C,Tr_\Q e^{i\Phi-in \oint A})(U)
$$
We can compute this by replacing $\Phi$ by
$\Phi'= \Phi- n \oint A$ everywhere. The only difference is that we must now
transform from $\Phi- n \oint A$ basis (with $\oint A$ as a momentum) where the computations are simple
to $\oint A$ basis with $\Phi$ as a momentum, and this is done by
$$
Z_{{\rm 2dYM}}(C)(U) = \int
d\Phi' \;e^{{1\over g_s} \Tr \Phi' u +{n\over 2 g_s} Tr u^2 }\; Z_{{\rm 2dYM}}(C)(\Phi').
$$
This gives
$$
Z(C,Tr_\Q e^{i\Phi- i n \oint A})(U) = \sum_{\R} q^{{n\over 2} C_2(\Q)}
S_{{\bar \Q} {\R}}  Tr_{\R} U
$$

\appendix{C}{Modular transformations}

\subsec{The vertex amplitude}
Consider the vertex amplitude corresponding to intersecting D4
branes:
$$
Z(U,V)=\sum_{\R \in U(N),\, \Q \in U(M)} {V}_{\R \Q} (N,M)
Tr_{\R} U Tr_{\Q}V \,\,
$$
where
$$
{V}_{\R \Q}=\sum_{\P \in U(M)}
q^{C_2^{(M)}(\P)\over 2}
S_{\R \P}(g_s,N)\, S_{\P \Q }(-g_s,M)
$$
Using the definition \defsr\ of $S_{\R\Q}$
and the Weyl-denominator form of the $U(N)$ characters
$Z(U,V)$ becomes:
$$
Z(U,V)={1\over \Delta_H(u) \Delta_H(v)} \sum_{w_1,w_3 \in S_N}
\sum_{w_3',w_2 \in S_M} (-)^{w_1+w_2+w_3+w_3'}
q^{\vert\vert \P +\rho_M \vert\vert^2 \over 2} $$
$$ q^{(\P+\rho_M)\cdot w_3'( \Q +\rho_M)}
q^{-(\P+\rho_N)\cdot \omega_3(\R +\rho_N)}
e^{i w_1( \R +\rho_N)\cdot u} e^{(\Q +\rho_M)\cdot w_2(i v)}
$$
We can trade the sums over the Weyl groups, for sums over the full weight
lattices: Put
$$w_2=w_Q^{-1},\quad w_3'=w_P^{-1} w_Q,
$$
this defines elements $w_P,w_Q \in S_M$ uniquely given $w_2,w_3'$.
Then, we can always find an element $w_R\in S_N$ such that
$$
w_3= w_P^{-1}w_R,
$$
for a given $w_3$, by simply viewing $w_P$ as an element of $S_N$
acting on first $M$ entries of any $N$ dimensional vector, leaving the others
fixed. Finaly, find an $w\in S_N$ such that
$$
w_1=w^{-1}w_R,
$$
Note now that
$$
\omega_P(P+\rho_N) = \omega_P (P+\rho_M) + \rho_N-\rho_M
$$
since $\omega_P$ acts only on first $M$ entries of a vector and
the first $M$ entries of $\rho_N-\rho_M$ are all equal, hence invariant under $\omega_P$.
Using this and the fact that
now only permutations $w$ are counted with alternating signs, we can
combine the sums ofer the weyl-groups with the sums over the lattices to
write:
$${
Z(U,V)=\Delta_H(u)^{-1} \Delta_H(v)^{-1} \sum_{w \in S_N}(-)^w
\sum_{m,p \in {\bf Z}^M; \;n \in {\bf Z}^N}
q^{p^2\over 2} q^{p\cdot m} q^{-(p+\rho_N-\rho_M)\cdot n}
e^{i n\cdot w(u)} e^{i m\cdot v} }
$$
Now split $n = (n',n'')$ where $n'$ is the first $M$ entries in $n$,
$n''$ the remaining $N-M$, and similarly put
$\rho_N-\rho_M = (\rho',\rho'')$,
where we have treated $\rho_M$ as $N$ dinemsional vector first
$M$ entries of which is the standard Weyl vector of $U(M)$, the remaining
being zero, and $u= (u',u'')$. If one in addition defines $m' = m-n'$
above becomes
%
%
%
%
%
$$\eqalign{
Z(U,V)= \theta^M(q)\Delta_H(u)^{-1} \Delta_H(v)^{-1} &\sum_{w \in S_N}(-)^w
\sum_{m'\in {\bf Z}^M} q^{-(m')^2\over 2} e^{i m'\cdot v} \cr
&\sum_{n' \in {\bf Z}^M} \sum_{n'' \in {\bf Z}^{N-M}}
q^{- \rho' \cdot n'- \rho'' \cdot n''}
e^{n'(\cdot w(i u')+i v) + n''\cdot \omega(i u'')} }
$$
where $\theta(q) = \sum_{n\in {\bf Z}} q^{n^2\over 2}$ is the usual theta function.
We write $n$ again as an $N-$dimensional vector
$(n',n') = n$ to get our final expression
$$
Z(U,V)=\theta(q)^{M} \Delta_H(u)^{-1} \Delta_H(v)^{-1}
\sum_{m'\in {\bf Z}^M} q^{-{(m')^2\over 2}} e^{i m'\cdot v}
\sum_{w \in S_N}(-)^w \delta(i v+w(iu)+ (\rho_N -\rho_M)g_s)
$$
where $v,\rho_M$ are regarded as $N$ dimensional vectors
$(v,0^{N-M})$, $(\rho_M,0^{N-M})$.

\appendix{D}{Large $N$ limit of the vertex amplitude}
Here we find the large $N,$ $M$ limit of the interaction
$$
{V}_{{\cal R} {\cal Q}}=\sum_{\P}S_{\R\P}(N,g_s)\, q^{C_2^{(M)}(\P)\over 2} \, S_{{\bar{\P}} \Q }(M,g_s)
$$
(we've dropped an overall factor).
Using $TS^{-1} = \theta(q)^M S^{-1}T^{-1}S^{-1}T^{-1}$ in the $U(M)$ factor, this can be done by
computing first the large $N,$ $M$ limit of
$$
\sum_{\P}S_{\R\P}(N,g_s)\, S_{{\bar{\P}} \A }(M,g_s)
$$
and then using large $M$ limit of $(TST)^{-1}_{\A \Q}$ to get the full amplitude.
In general, either version of the problem is very difficult and at present unsolved. Things simplify significantly if we $turn$ $off$ the $U(1)$ charges
all together. This means we will effectively compute the $SU(N)$ rather than
$U(N)$ version of interaction. It will turn out that the crucial features that one expects from the amplitudes $assuming$ the conjecture holds,
are unaffected by this. In this case, the representations $\R$ are effectively
labeled by Young tableaux's.

From the free fermion description of the $YM$ amplitudes
it follows easily \AOO\ that:
\eqn\limitnew{\eqalign{
\sum_{{\P}\in U(M)}
S_{{\R} {\P}}(g_s,N) & S_{{\bar \P} {\A}}(g_s,M)
 \rightarrow
\; \alpha^{-1}_N(q) \alpha^{-1}_M(q) \; S_{0 (R_+ {\bar R_-})} (g_s,N)  S_{0 (A_+ {\bar A_-})} (g_s,M)\cr
& \times
\prod_{i,j=1}^{\infty}
{[\half N - \half M +j-i] \over [{R}^+_i-{A}^+_j+\half N - \half M+j-i] } \,\,
 {[\half N -\half M +j-i] \over [{R}^-_i-{A}^-_j+\half N-\half M+j-i] }\cr
& \times
\prod_{i,j=1}^{\infty}
 {[\half N+\half M -j-i+1] \over [{R}^+_i+{A}^-_j+\half N+\half M -j-i+1] } \, \,
{[\half N +\half M -j-i+1] \over [{R}^-_i+{A}^+_j+\half N +\half M-j-i+1] }}
}
where the arrow indicates taking large $N,\,M$ limit and where
$
\alpha_N(q) = q^{-(\rho_N^2 +{N\over 24})} M(q)\eta^N(q),
$
and similarly for $\alpha_M$ with $M,N$ exchanged.

For simplicity, we will be are interested only in the leading chiral block
of the amplitude which determines the Calabi-Yau manifold that the YM theory describes in the large $N$ limit, and neglects the excitations of
non-normalizable modes.
In this limit, the piece $S_{0 (R_+ {\bar R_-})} (g_s)  S_{0 (A_+ {\bar A_-})} (g_s)$ gives
$$\alpha_M(q) \alpha_{N}(q)  W_{A_+^T 0}(q) W_{A_-^T 0}(q)  W_{R_+^T 0}(q) W_{R_-^T 0}(q)
q^{-{M(\vert A_+\vert +\vert A_-\vert)\over 2}}q^{-{N(\vert R_+\vert +\vert R_-\vert)\over 2}}
$$
where $W_{RP}(q)=s_R(q^{\rho})s_P(q^{R+\rho}),$
and moreover $W_{R0}(q)=(-)^{|R|}q^{k_{R}/2} W_{R^T0}(q).$
Of the infinite product terms, in the leading chiral block limit
only the second row in \limitnew\ contributes. This is because
the interactions between the chiral and anti-chiral part of the amplitude are
supressed in this limit. Using
$$
\prod_{i,j}{1\over x_i-y_j}=\prod_{i}x_i^{-1}\,\sum_{R} s_R(x^{-1})s_R(y)
$$
we get
$$ const. \times\sum_{P_+,P_-} s_{P_+} (q^{R_+ +\rho}) s_{P_+} (q^{-(A_+ + \rho)})
s_{P_-} (q^{R_- + \rho})
s_{P_-} (q^{-(A_- +\rho)})
q^{(\vert P_+ \vert +\vert P_- \vert) {N-M\over 2}}
$$
The constant comes from regularizing the infinite products
(see \AOO\ for details) and can be determined by computing the leading large $N$, $M$ scaling
$$
\eqalign{ &\prod_{(i,j)\in A_+} {[\half N-\half M-j+i] \over [-\half N-\half M -j+i]}\,\,
\prod_{(i,j)\in A_-} {[\half N-\half M -j+i] \over
[-\half N-\half M-j+i]}\,\, \cr
&\prod_{(i,j)\in R_+} {[\half N-\half M +j-i] \over [\half N+\half M+j-i]}\,\,
\prod_{(i,j)\in R_-} {[\half N-\half M +j-i] \over [\half N+\half M +j-i]}
\sim
q^{\kappa_{A_+} +\kappa_{A_-} \over 2} \,\, q^{M (\vert R_+ \vert +\vert R_-\vert+\vert A_+ \vert +\vert A_-\vert )\over 2} }
$$
where $i$ goes over the rows and $j$ over the columns.
All together, this gives
$$\eqalign{\sum_{{\P}\in U(M)}
S_{{\R} {\P}}(g_s,N) S_{{\bar \P} {\A}}(g_s,M) \rightarrow&
 (-)^{\vert R_+\vert+\vert R_-\vert }
q^{-{N-M\over 2}(\vert R_+\vert + \vert R_-\vert)}
q^{-{\kappa_{R_+} +\kappa_{R_-}\over 2}} q^{\kappa_{A_+} +\kappa_{A_-} \over 2}
\cr
&\sum_{P_+} (-)^{\vert P_+\vert}
W_{R_+ P_+}(q) W_{P_+^TA_+^T} (q)q^{{N-M\over 2}\vert P_+ \vert }\cr
&
\sum_{P_-} (-)^{\vert P_-\vert}
W_{R_- P_-} (q) W_{P_-^TA_-^T} (q)
q^{{N-M\over 2}\vert P_- \vert}
}
$$
Next, recall that (see appendix A) the large $M$ limit (more precisely, the leading chiral block) of $(TST)^{-1}$ is
$$
(T^{-1}S^{-1}T^{-1})_{AQ} = \alpha_M(q^{-1})
W_{A_+ Q_+}(q) W_{A_-Q_-} (q)
q^{-{\kappa_{A_+} +\kappa_{A_-}+ \kappa_{Q_+} +\kappa_{Q_-}\over 2}}
$$
To compute our final expression, we need to sum:
$$\eqalign{
\sum_{\P}S_{\R\P}(N,g_s)\,
q^{C_2^{(M)}(\P)\over 2} \,& S_{{\bar{\P}} \Q }(M,g_s) \rightarrow 
\alpha_M(q^{-1}) (-)^{\vert R_+\vert+\vert R_-\vert }
q^{-{N-M\over 2}(\vert R_+\vert + \vert R_-\vert)}\cr
& q^{-{\kappa_{R_+} +\kappa_{Q_+}\over 2}} 
\sum_{P_+, A_+} (-)^{\vert P_+\vert}
W_{R_+ P_+}(q) W_{P_+^TA_+^T} (q) W_{A_+Q_+}(q)q^{{N-M\over 2}\vert P_+ \vert }\cr
&
q^{-{\kappa_{R_-}+ \kappa_{Q_-}\over 2}}\sum_{P_-,A_-} (-)^{\vert P_-\vert}
W_{R_- P_-} (q) W_{P_-^TA_-^T} (q) W_{A_-Q_-}(q)
q^{{N-M\over 2}\vert P_- \vert}.
}
$$
Note that this contains an
ill-defined expression
\eqn\delid{
\sum_{A_+} {W}_{P^T_+ A^T_+}(q)\, {W}_{Q_+ A_+}(q)\,\,
\sum_{A_-} {W}_{P^T_- A^T_-}(q)\, {W}_{Q_- A_-}(q)
}
The physical interpretation of a finite version of this amplitude,
with insertions of $e^{-t|A_+|}$ and $e^{-{\bar t}|A_-|}$
also suggests how to define \delid . 
Namely, the finite amplitude is the
topological string amplitude (more precisely, two copies of it)
on ${\cal O}(-1)\oplus {\cal O}(-1) \rightarrow
{\rm\bf P}^1$ with D-branes as in the figure 6, where the size of the
${\rm\bf P}^1$ is $t$. 
In the limit $t\rightarrow 0$ the ${\rm\bf P}^1$
shrinks to zero size, and one can undergo a conifold transition, to a small $S^3$ of size $\epsilon$. In this case, the only holomorphic maps correspond to those with $P_+=Q_+$, so that
$$
\sum_{A_+} {W}_{P^T_+ A^T_+}(q)\, {W}_{Q_+ A_+}(q)\,\,
=\delta(P_+ - Q_+),
$$
and similarlty for the anti-chiral piece,
which is independent of $\epsilon$, as this is a
complex structure parameter.

\bigskip
\centerline{\epsfxsize 3.0truein\epsfbox{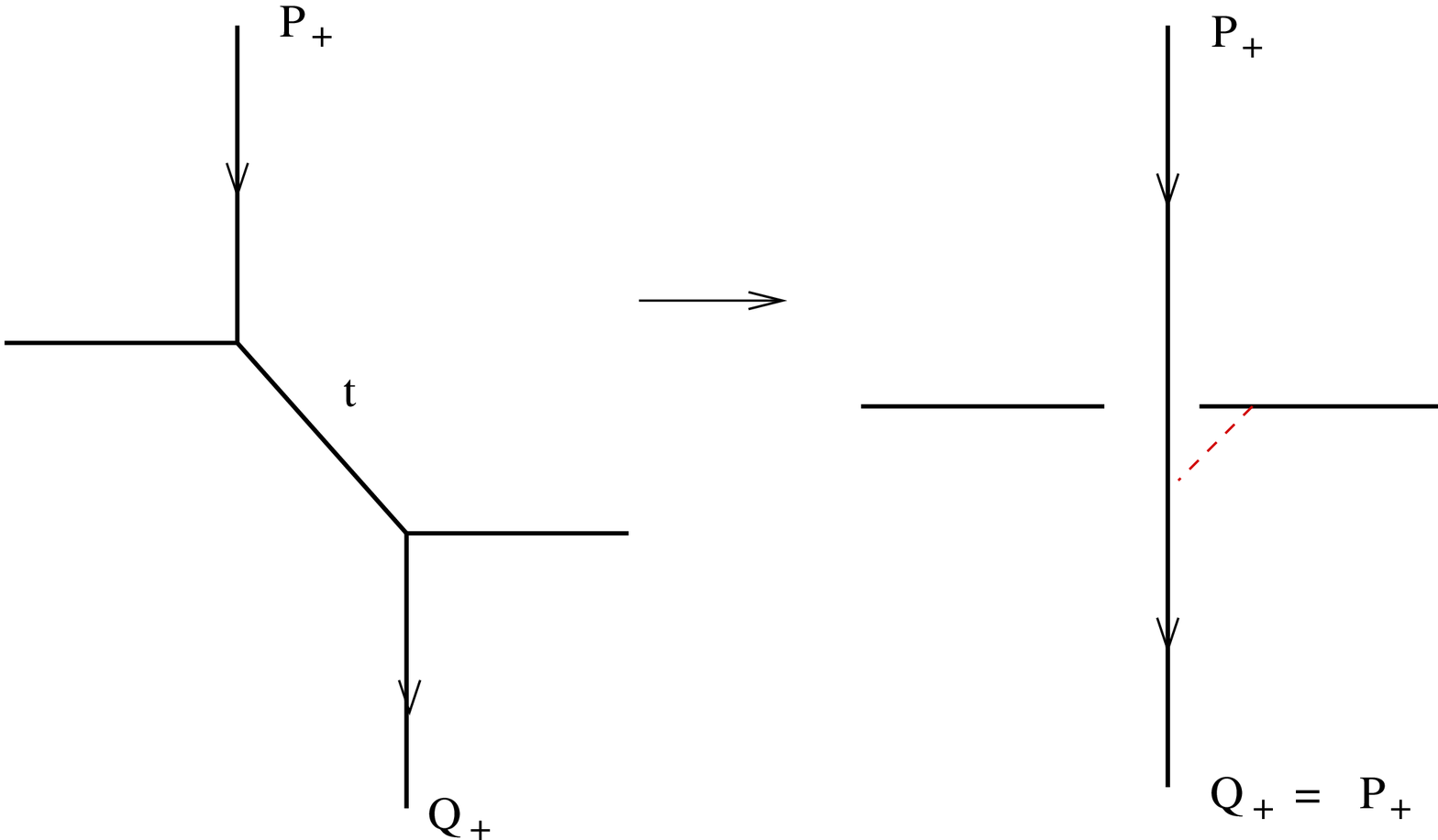}}
\leftskip 2pc
\rightskip 2pc \noindent{\ninepoint\sl
\baselineskip=2pt {\bf Fig. 6.}
{{The figure on the left corresponds to $O(-1)\oplus O(-1)\rightarrow
{\rm\bf P}^1$ with ${\rm\bf P}^1$ of size $t$ with two stacks of lagrangian D-branes.
The representations $P_+$ and $Q_+$ label the boundary conditions on open string maps. When $t=0$ the Calabi-Yau is singular, but can be desingularized by growing a small $S^3$. The singular topological string amplitudes can be regulated correspondingly, and with this regulator, they vanish unless $P_+=Q_+$. See \AMV\ for more details.}}}
\bigskip

Our final result is:
\eqn\final{\eqalign{
\sum_{\P}S_{\R\P}&(N,g_s)\,
q^{C_2^{(M)}(\P)\over 2} \, S_{{\bar{\P}} \Q }(M,g_s) \rightarrow
 \alpha_M(q^{-1})\, \theta^M(q)\,\, (-)^{\vert R_+\vert+\vert R_-\vert +\vert Q_+\vert +
\vert Q_-\vert}\cr
&q^{-{N-M\over 2}(\vert R_+\vert + \vert R_-\vert)}
q^{{N-M\over 2}(\vert Q_+ \vert+\vert Q_- \vert)}
q^{-{\kappa_{R_+} +\kappa_{R_-}\over 2}}
q^{-{\kappa_{Q_+}+ \kappa_{Q_-}\over 2}}
W_{R_+ Q_+}(q)
W_{R_- Q_-} (q) }}

\listrefs

\bye